\documentclass[pdflatex,sn-mathphys-num]{sn-jnl}

\usepackage{graphicx}%
\usepackage{multirow}%
\usepackage{amsmath,amssymb,amsfonts}%
\usepackage{amsthm}%
\usepackage{mathrsfs}%
\usepackage[title]{appendix}%
\usepackage{xcolor}%
\usepackage{textcomp}%
\usepackage{manyfoot}%
\usepackage{booktabs}%
\usepackage{algorithm}%
\usepackage{algorithmicx}%
\usepackage{algpseudocode}%
\usepackage{listings}%
\usepackage[utf8]{inputenc}
\usepackage{graphicx}
\usepackage{quoting}
\usepackage{geometry}
\usepackage{enumitem}
\usepackage{float}
\usepackage[labelfont=bf]{caption}
\usepackage{subcaption}
\usepackage{setspace}
\usepackage{multirow}
\usepackage{geometry} 
\usepackage[maxbibnames=99, style=nature]{biblatex} 
\usepackage{array}
\newcolumntype{P}[1]{>{\centering\arraybackslash}p{#1}}
\newcolumntype{M}[1]{>{\centering\arraybackslash}m{#1}}

\usepackage{xr}
\newcommand*{\addFileDependency}[1]{
\typeout{(#1)}
\IfFileExists{#1}{}{\typeout{No file #1.}}
}\makeatother

\newcommand*{\myexternaldocument}[1]{%
\externaldocument{#1}%
\addFileDependency{#1.tex}%
\addFileDependency{#1.aux}%
}
\myexternaldocument{supplementary}

\begin{document}

\title{Are EVs Cleaner Than We Think? Evaluating Consequential Greenhouse Gas Emissions from EV Charging}

\author[1]{\fnm{Riti} \sur{Bhandarkar}}\email{ritib@princeton.edu}
\equalcont{These authors contributed equally to this work.}
\author*[1]{\fnm{Qian} \sur{Luo}}\email{ql7299@princeton.edu}
\equalcont{These authors contributed equally to this work.}
\author[1]{\fnm{Emil} \sur{Dimanchev}}\email{ed0400@princeton.edu}
 
\author[1,2]{\fnm{Jesse} \sur{D. Jenkins}}\email{jdj2@princeton.edu}

\affil[1]{\orgdiv{Andlinger Center for Energy and the Environment}, \orgname{Princeton University}, \orgaddress{\city{Princeton}, \postcode{08540}, \state{NJ}, \country{USA}}}
\affil[2]{\orgdiv{Department of Mechanical and Aerospace Engineering }, \orgname{Princeton University}, \orgaddress{ \city{Princeton}, \postcode{08540}, \state{NJ}, \country{USA}}}


\abstract{While electrifying transportation eliminates tailpipe greenhouse gas (GHG) emissions,  electric vehicle (EV) adoption can create additional electricity sector emissions. To quantify this emissions impact, prior work typically employs short-run marginal emissions or average emissions rates calculated from historical data or power systems models that do not consider changes in installed capacity. In this work, we use an electricity system capacity expansion model to consider the full consequential GHG emissions impact from large-scale EV adoption in the western United States, accounting for induced changes in generation and storage capacity. We find that the metrics described above do not accurately reflect the true emissions impact of EV adoption--average emissions rates can either \textit{under}- or \textit{over}-estimate emission impacts, and short-run marginal emissions rates can significantly \textit{underestimate} emission reductions, especially when charging timing is flexible. Our results also show that using short-run marginal emission rates as signals to coordinate EV charging could increase emissions relative to price-based charging signals, indicating the need for alternative control strategies to minimize consequential emissions.
}

\maketitle
\section{Introduction}
More than 17 million electric vehicles (EV) and plug-in hybrid electric vehicles (PHEVs) were sold worldwide in 2024, accounting for more than one in five new light-duty vehicles\cite{ev_sales}. EV adoption is expected to increase in the coming years as the cost and performance of electric vehicles continue to improve and policies support the phaseout of internal combustion engine vehicles (ICEVs) and reduction of greenhouse gas emissions (GHGs)\cite{evo_bnef, evo_iea, ev_emissions_epa}. 

While electrifying transportation eliminates tailpipe GHG emissions, the emissions impact of EV adoption also depends on any induced changes in electricity sector emissions. To elucidate trade-offs and benefits associated with EV adoption, a substantial body of prior work estimates induced power sector emissions or net `well-to-wheels' emissions associated with EV adoption across varying geographies and contexts\cite{choi_2018, chen_2021, gagnon_planning_2022, vega2023mapping, wu2024adoption, sharma2024multisectoral, singh2024ensuring}. Prior work typically employs one of two metrics to estimate power sector emissions associated with EV charging: average emissions rates or short-run marginal emissions rates (see Table \ref{tab:emisMetrics}). For example, a variety of studies estimate the `well-to-wheels' emissions intensity for EVs in different regions across the world using average emissions rates for historical or projected future grids \cite{qiao2019life, andersson2021greenhouse, vega2023mapping, wu2024adoption}. Some studies have compared the emission impact when average and marginal emission factors are used in the analysis and found significant differences when the two metrics are used\cite{tamayao_regional_2015, zhong2023revisiting, }. While employing marginal emissions rates, some studies conclude that EV adoption may result in higher GHG emissions than equivalent ICEVs or gasoline hybrids, depending on the location and brand of EVs\cite{vega2023mapping, singh2024ensuring}.

\begin{table}[ht]
\centering
\caption{\label{tab:emisMetrics}Taxonomy of emissions impact metrics}
\begin{tabular}{P{0.25\textwidth}P{0.65\textwidth}}
\hline
Metrics & Definition\\
\hline
Average emissions rate (AER) & Total emissions divided by total electricity supply or consumption.\\
\hline
Short-run marginal emissions rate (SR-MER) & The emissions change per unit change in electricity consumption, holding the structure of the electrical system (e.g., the generation, storage and network capacity) fixed. \\
\hline
Long-run marginal emissions rate (LR-MER) & The emissions change per unit change in electricity consumption, accounting for induced changes in both the operation and structure (installed capacity) of the electricity system.\\
\hline
Consequential emissions & The change in emissions caused by a specific intervention, action, or change in supply or demand.\\ 
\hline
\end{tabular}
\end{table}

However, it is unlikely that either average or short-run marginal emissions rates reflect the true consequential emissions impact of large-scale adoption of electric vehicles. Average emissions rates reflect the total generation mix of a given region and thus primarily reflect historical capacity decisions while disregarding how that mix will evolve in specific response to changes in demand. Likewise, short-run marginal emissions rates estimate marginal changes in power system emissions in response to small (marginal) changes in electricity demand, accounting only for operational changes in dispatch while holding the structure of the power system constant. Crucially, both metrics neglect how large-scale changes in electricity demand and/or supply affect the structural evolution of the power system, including deployment and retirement of electricity generation, storage, or network capacity.

Prior work has demonstrated that omitting these potential structural changes can significantly impact estimates of consequential GHG emissions. Hawkes (2014) found that analyses based on short-run marginal emissions significantly overestimated the emissions impacts of heat pump deployment in the UK\cite{hawkes_long-run_2014}. Gagnon and Cole (2022) estimated consequential GHG emissions for a set of 31 different electricity load (demand) shapes and likewise found that long-run marginal emissions accounting for structural changes were substantially lower than short-run marginal emissions rates, while using average emissions rate could both over- and under-estimate total consequential emissions in different cases\cite{gagnon_planning_2022}. Holland et al. (2022) also conclude that long-run emissions “differ in surprising ways from their short-run analogs.”\cite{holland2022decarbonization} Finally, recent work assessing the consequential emissions impact of various clean electricity procurement strategies to meet the demand of new hydrogen electrolysers \cite{ricks_minimizing_2023, giovanniello2024influence} and corporate \& industrial demand \cite{xu2024system, riepin2024means} found substantial differences between expected impacts of short-run marginal emissions-based procurement strategies and the actual consequential impacts when accounting for structural changes.

Estimating the consequential emissions impacts of EV adoption is also complicated by the fact that the timing of EV charging may be highly flexible. For example, prior work finds that controlled charging that redistributes charging demand throughout the night during periods of lower average emissions may result in higher short-run marginal emissions in some U.S. regions, as coal-fired generators are often on the operating margin at night \cite{jochem_assessing_2015,tamayao_regional_2015,mclaren_emissions_2016}. Other work finds that shifting demand to day-time hours when solar PV production is curtailed reduces short-run emissions \cite{arvesen_emissions_2021,powell_charging_2022,lauvergne_integration_2022}. Furthermore, many studies have recommended rescheduling EV charging based specifically on short-run marginal emission rates, rather than prices, to maximize emission reductions \cite{dixon_scheduling_2020, tu_electric_2020, huber_carbon_2021, pepiciello2022real, , chen2022emission, wang2022feasibility, kang2023sustainable, li_smart_2023}. One recent study attempts to incorporate some structural change in the grid by optimizing EV charging based on planned investments over a 5-year horizon, using what they term a "medium-run marginal emissions factor"\cite{powell2024future}.  However, this approach still assumes that there is no dynamic interaction between EV demand and grid investments. Commercial products, like the Automated Emissions Reduction tools offered by WattTime, are now available to implement emissions-based charging schedules\cite{watttime}. However, each of these controlled charging strategies also neglect potential structural effects. At present, the literature is unclear about how controlled charging based on short-run marginal emissions rates impacts capacity entry and exit decisions in the electricity sector. While the charging behavior of a single vehicle is too small to affect capacity decisions, investors in power systems assets certainly attempt to anticipate large-scale changes in both demand and competing supply and adjust their decisions accordingly. Given that prior literature indicates substantial divergences between short- and long-run estimates of consequential emissions in other contexts, we should likewise expect that aggregate changes in demand due to widespread use of one charging control strategy or another can have substantially different consequential impacts. Unfortunately, the true consequential impact of large-scale changes in demand or supply, accounting for both operational and structural effects, is empirically unobservable and can only be approximated by modeling counterfactual scenarios. It is thus unclear what practical control signals and strategies are best suited to coordinate EV charging (or other flexible demand) in order to minimize consequential emissions.

In this work, we use an open-source electricity system capacity expansion model, GenX\cite{jenkins2017enhanced, genx} to consider the full consequential GHG emissions impact from large-scale light-duty EV adoption in the western United States (WECC, see figure S1) circa 2030, including induced changes in transmission network expansion, installed generation and storage capacity, and associated operations. We also compare these estimates to emissions impacts attributed based on short-run marginal and average emissions rates. This study offers insights into the GHG emissions impact of transportation electrification and the utility of alternative metrics for consequential emissions impacts. First, we demonstrate that using capacity expansion modeling to account for both structural and operational changes provides a more complete assessment of expected consequential emissions impacts from EV adoption than either short-run marginal or average emissions rate metrics. Second, we find that short-run marginal emissions tend to significantly overestimate electricity sector emissions associated with EV adoption and thus \textit{underestimate} emissions benefits from vehicle electrification; similarly, we also find that using average emissions rates can overestimate emissions caused by EV adoption in most regions in WECC. Third, we show that scheduling EV charging to avoid short-run marginal emissions could increase emissions relative to price-based charging signals, indicating the need for alternative control strategies to minimize consequential emissions.

\section{Short-run metrics strongly overestimate and average emission rates can both over and underestimate emissions from EV adoption} 

To estimate the consequential emissions impact of EV charging demand across the WECC, we first assume on-road EV penetration in 2035 reaches 15.6 million EVs in the region (approx. 20\% penetration),  consistent with the REPEAT Project's mid-range estimate of EV adoption in 2030 under current U.S. federal and state policies \cite{repeat}. We then perturb EV penetration and resultant charging demand by +5\% intervals at both the WECC-wide level as a whole and for each of the modeled zones one at a time. At the WECC-wide level, 5\% perturbation in EV adoption constitutes an increase of about 780,000 EVs, which adds about 10 MW of demand during the lowest hour in the EV load profile and 900 MW during peak coincident demand. We then calculate the full consequential changes in WECC-wide emissions, inclusive of effects on capacity additions/retirements (``long-run marginal emissions''), emissions attributed based on short-run marginal emissions rates, and average emissions rates. See Methods for further detail on experimental design and calculation of emissions attributed to EV charging under each metric.

\subsection{Long-run  v.s. short-run} 
\begin{figure}[h]
\centering
\includegraphics[width=1.0\textwidth]{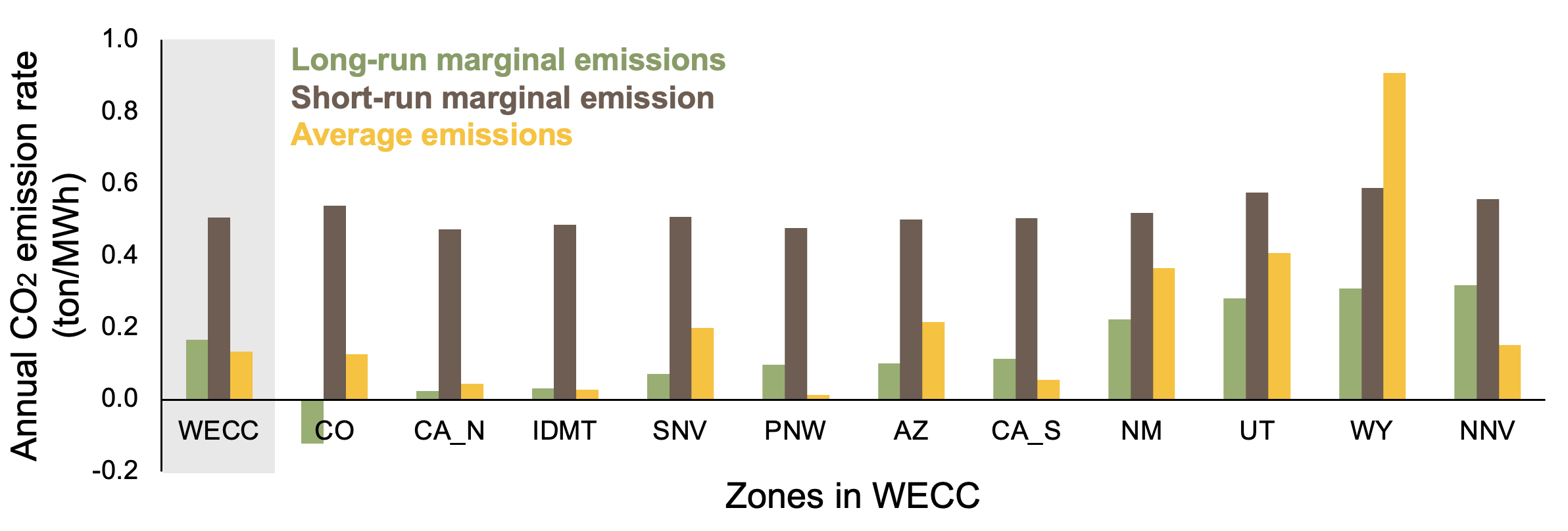}
\caption{Annual long-run marginal, short-run marginal, and average power system emission rates in WECC. For each zone, the marginal emission rates are calculated for WECC as a whole and based on a 5\% EV adoption rate increase in each zone; the average emission rates are calculated for each zone based on the zonal emissions and demand. Results are based on the ``Base" EV stock numbers (15.6 million EVs).}
\label{emission}
\end{figure}

\begin{figure}[h]
\centering
\includegraphics[width=1.0\textwidth]{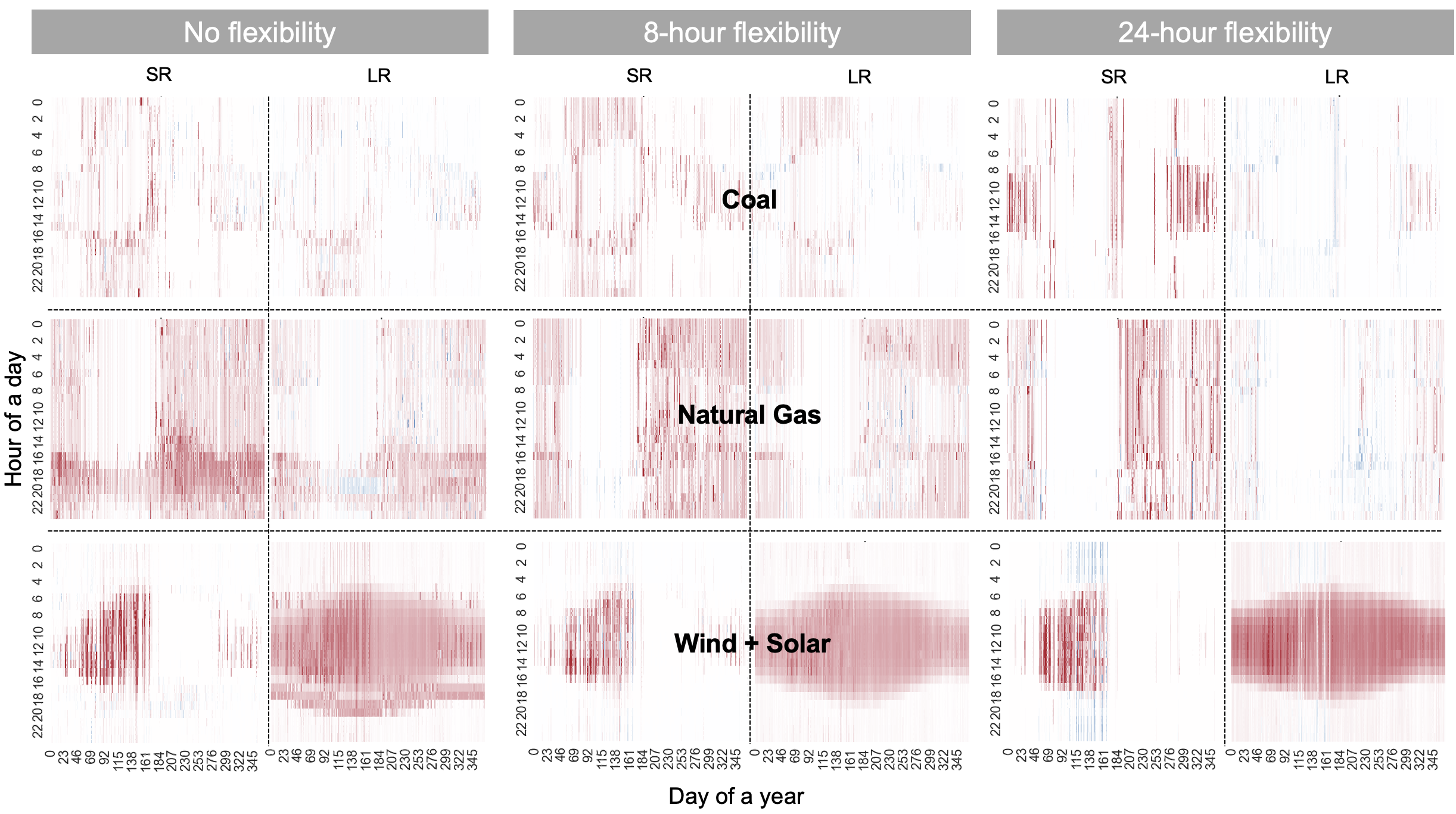}
\caption{Short-run (``SR'') and long-run (``LR'') marginal generation profiles for different technologies due to 5\% EV demand perturbation under various charging flexibility levels for mid-electrification baseline case (15.6 million EVs). Positive (red) values indicate that the generation increased in that hour when demand was added, while negative (blue) values indicate that the generation was decreased in that hour when demand was added.}
\label{margGenHr}
\end{figure}

Across all tested EV adoption levels and regions, we find that the estimated emissions intensity of EV charging is substantially lower when based on long-run marginal emissions than if short-run marginal emission rates are employed (Figures \ref{emission} and S2). Using short-run marginal emissions metrics, EV charging in WECC is responsible for about 0.47-0.59 tCO$_2$/MWh across each zone, reflecting the typical role of emitting coal and natural gas power plants as the marginal generators in each region during most hours of charging demand (``No flexibility'' case in Figure \ref{margGenHr}). In contrast, long-run marginal emissions accounting for structural changes in capacity induced by EV adoption results in estimated emissions intensity of 0.17 tCO$_2$/MWh WECC-wide and -0.12 to +0.32 tCO$_2$/MWh across zones. Long-run emissions rates are considerably lower because they account for additional deployment of solar, wind and battery storage induced by increased demand from EV charging (Figure S3). In the context of current federal and state policies and given technology and fuel costs, the composition of new capacity additions is considerably cleaner than the marginal generators in the region. 

To explore how estimates of emissions intensity change at varying EV penetration levels, we increase EV penetration in 5\% intervals across +/- 15\% variation from the base EV stock assumptions and recalculate emissions intensity for each increment of demand. The short-run marginal emission rates at the WECC level are consistently around 0.51 tCO$_2$/MWh due to the static power system. The long-run marginal emission rates range between 0.15 and 0.18 tCO$_2$/MWh, 64-70\% lower than the equivalent short-run estimate (Figure S2 (a)). In addition to the EV adoption level, we also conduct a sensitivity analysis covering variations in wind, solar and battery investment costs, natural gas prices, and the degree of flexibility in EV charge schedules. Across sensitivities, short-run marginal emission rates are 0.28-0.57 tCO$_2$/MWh greater than the long-run marginal emission rates (which span -0.02 to 0.17 tCO$_2$/MWh across all scenarios, Figure S2 (b)). We present a comprehensive analysis of the impacts of EV charging flexibility in Section \ref{sec:flexibility}.

Compared with the average annual emissions from an internal combustion engine vehicle (ICEV, approx. 3 tCO\textsubscript{2}/year), emissions associated with an EV are much lower in WECC regardless of which metric is applied. However, the estimated emissions reductions benefits from switching from ICEV to EV are significantly underestimated by short-run marginal emissions metrics. Using long-run marginal emissions estimates, substituting an average EV for an average ICEV reduces CO$_2$ emissions from the fuel cycle by 67-86\% depending on the EV adoption level, while short-run emissions metrics would attribute only a 35-39\% reduction (Figure S2 (a)).

The discrepancy between the long run and short run emissions intensity is highly dependent on current and future generation mix, and each state served by WECC is characterized by different infrastructure and demand (e.g., Montana has a generation mix dominated by coal, while the majority of Washington's electricity comes from hydroelectricity). As such, we also estimate system-wide consequential emissions while perturbing EV adoption only in a single zone, holding demand in the other 10 zones constant at the base adoption level. Here, we find that long-run marginal emission rates are always at least 40\% lower than short-run emission rates when EVs are added to different zones in WECC (Figure \ref{emission}). While the short-run marginal emission intensity is very consistent across different zones (0.47-0.59 tCO$_2$/MWh), long-run emission intensity estimates vary significantly driven by regional differences in structural capacity effects. When 5\% additional EVs (155 MW aggregate peak charging power and 360 GWh annual charging demand) are adopted in Colorado (``CO''), the model builds 41 MW of wind and 103 MW of solar and retires 19 MW of natural gas in the WECC system. As a result of combined structural and operational changes, every TWh of EV demand added to CO reduces coal-fired generation by 61 GWh and gas-fired generation by 122 GWh, resulting in a negative system wide emission rate of -0.12 tCO$_2$/MWh. As a contrasting example, EV adoption in North Nevada (``N. NV'') results in a system-wide emissions rate of 0.32 tCO$_2$/MWh. In this zone, a 1 TWh increase in EV charging demand results in structural and operational changes that drive a 122 GWh increase in coal-fired generation and 348 GWh increase in natural gas-fired generation, with the remainder of demand met by renewable energy (Figure S4). While accounting for structural capacity changes results in more variation in long-run emissions intensity estimates than short-run estimates, we find that long-run emissions intensity of EV charging is consistently lower than short-run metrics would indicate across all WECC regions (43\% to 123\% lower across zones).

\subsection{Long-run vs. average emissions rates} 

Average emissions rates are also frequently used in lifecycle assessment and well-to-wheels fuel cycle emissions intensity estimates. In the case of EV adoption in the Western US, we find the average emissions rate in WECC is 0.13 tCO\textsubscript{2}/MWh, close to the long-run marginal rate. This result agrees with the finding in Gagnon et al. that the average emissions rate can do a better job than the short-run emissions rate at approximating the long-run emissions impact of an intervention in grids with clean generation\cite{gagnon_planning_2022}. However, this is largely coincidental. Average emissions rates largely reflect the historical installed capacity. Thus in the case of the WECC as a whole, the average emissions rate includes contributions from existing inframarginal renewable, hydropower and nuclear generators and thus is lower than marginal emissions rates (where marginal generation in the region is most frequently a fossil fuel-burning power plant). In other contexts, however, it is possible for the average emissions intensity to be quite high (e.g., in a coal-dominated grid region), while the emissions intensity of new-build resources induced by increased demand is low (e.g., primarily consisting of wind and/or solar PV). To illustrate this dynamic, we estimate emissions attributed to EV charging using average emissions rates in each of the WECC zones when EV demand is increased only in that specific zone. We find that using average emission rates, the estimated emissions intensity of EV charging is substantially higher than long-run marginal emissions in six of eleven model zones, substantially lower in three, and roughly equal in only two (Figure \ref{emission}).


\section{Demand flexibility of EV charging unlocks the potential for greater emissions reductions that could be underestimated by the short run metric}\label{sec:flexibility}

The analysis above assumes a static EV demand profile, reflecting vehicle owners charging when most convenient without exposure to dynamic electricity prices. However, EV charging can be a highly flexible load, with the potential to shift the timing of charging considerably if provided appropriate incentives and/or controls. We thus explore the impact of two flexible charging scenarios on consequential emissions from EV charging: (1) the ability to delay charging demand by up to 8 hours and (2) a 24 hour flexibility case permitting EV charging to shift forward or backwards up to 12 hours. The former is consistent with common incentive programs to charge overnight during off-peak periods, while the latter would imply drivers had access to ubiquitous charging infrastructure both at home and during working hours and could thus choose the ideal time to charge during the full day. In both cases, charging demand is scheduled to minimize overall costs.

\begin{figure}[h]
\centering
\includegraphics[width=1.0\textwidth]{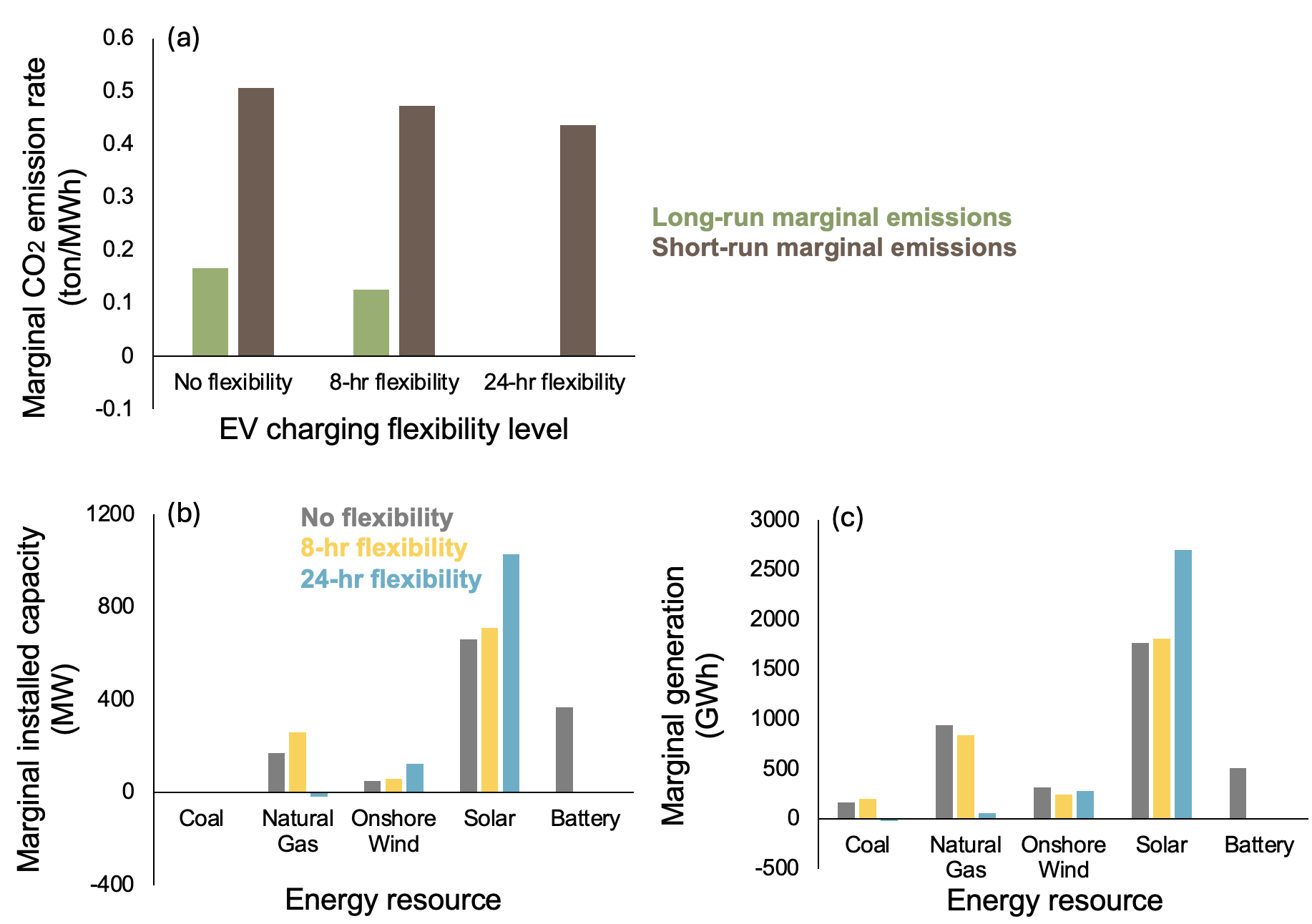}
\caption{WECC system-level marginal emission rates (a), marginal installed capacity (b), and marginal annual generation (c) of energy resources due to 5\% EV demand increase (780,000EVs) in WECC under different EV charging flexibility levels .}
\label{margEmisCapGen_flex}
\end{figure}

We find that the ability to delay EV charging demand by 8 hours or choose when to charge within a 24 hour period results in lower estimated emissions intensity using both long-run and short-run marginal emissions estimates(Figure \ref{margEmisCapGen_flex} (a)). However, the gap between short-run and long-run marginal emission rates is larger in these flexible charging scenarios, especially when the flexibility level is high (``24-hr flexibility'' case). 

\begin{figure}[h!]
\centering
\includegraphics[width=1.0\textwidth]{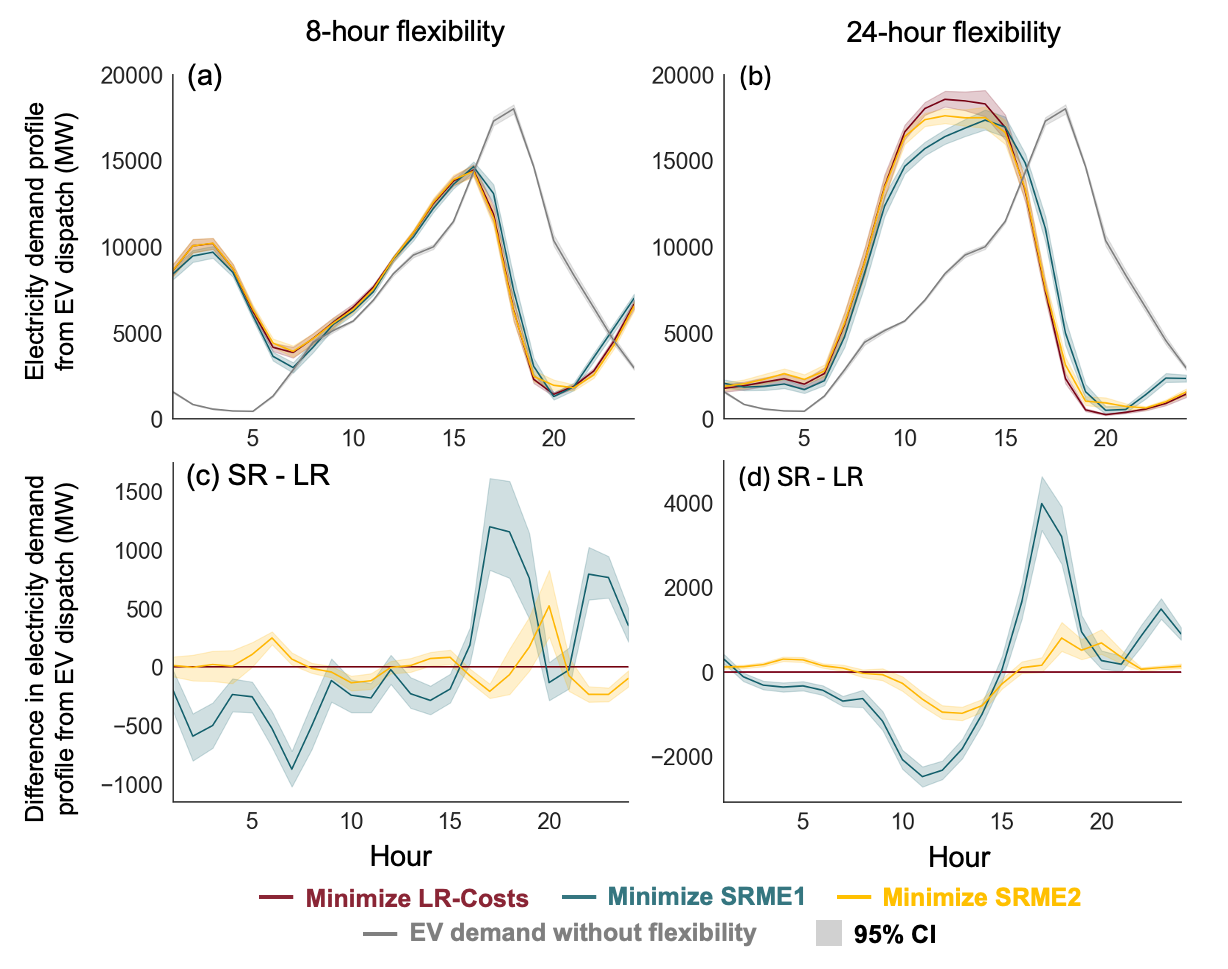}
\caption{EV charging demand profiles (a and b) and the charging demand differences between the short-run marginal emission minimization (``Minimize SRME1\&2'') and long-run cost minimization cases (``Minimize LR-Costs'') in WECC under 8- and 24-hour flexibility levels.}
\label{dispatch}
\end{figure}

When 8-hour charging flexibility is allowed, EV charging shifts from peak-load hours in evenings to night and early morning hours (Figure \ref{dispatch}). With fixed capacities (``SR'' case in Figure \ref{margGenHr}), this demand shift causes more marginal generation from coal and gas from late night to early morning, but marginal generation from fossil fuels significantly drops during the evening peak, making the short-run marginal emission intensity of EV charging lower than that under the ``No flexibility'' scenario. The long-run case also shows a similar pattern, resulting in a lower long-run marginal emission intensity. However, by shifting more demand to the night, instead of the daytime when solar resources are available, EV charging demand does not significantly promote the addition of renewable energy to the system compared with the ``No flexibility'' case. Instead, more natural gas capacity is added to meet the mid-night demand increase and more marginal generation is from coal-fired combustion (Figure \ref{margEmisCapGen_flex}). 

In the 24-hour flexibility case, a larger share of EV charging occurs during daytime periods, aligning with periods when low-cost solar power is abundant. Without any capacity changes (``SR'' case in Figure \ref{margGenHr}), shifting EV charging profiles causes some coal generation to shift to the daytime and gas generation is more evenly distributed across the day (e.g. less required in evening peak periods), but overall marginal generation from fossil fuels is reduced, leading to slightly lower short-run marginal emissions (Figure \ref{margEmisCapGen_flex}(a)). However, when considering the impacts of EV charging on installed capacity (``LR'' case), increased demand during daytime hours in the 24-hour flexibility case induces significant increases in solar PV capacity additions and decreases in natural gas capacity (Figure \ref{margEmisCapGen_flex}(b)). In this case, all incremental EV demand is met by renewable energy, as EV charging shifts to the lowest price periods, which also align with periods of abundant wind and solar availability, resulting in near-zero long-run marginal emissions (Figure \ref{margEmisCapGen_flex}(a)). This 24-hr flexibility result is consistent with a growing number of studies that demonstrate the potential for significant emissions reductions from charging at workplaces (or elsewhere) during the daytime \cite{powell_charging_2022, lauvergne_integration_2022, jochem_assessing_2015}. In our case study, the ``8-hr flexibility'' and ``24-hr flexibility'' cases also reduce system costs by \$600 and \$900 million compared with the ``No flexibility'' case, respectively, or \$10 to 15 per MWh of charging demand. 


\section{Re-scheduling EV charging to minimize short-run marginal emissions may not result in expected emission benefits in long term.}

While shifting EV charging to minimize costs tends to reduce emissions in systems where periods of abundant renewable and other low-carbon energy supply correlate with low prices (as in our case study above), it is theoretically possible to further reduce emissions by dispatching EV charging to minimize emissions instead of costs \cite{daniels2022more}. However, the variety of ways to estimate marginal emissions, including short- and long-run methods, and their diverging implications (demonstrated above) suggest that the effectiveness of such emissions-based charging may depend strongly on the chosen control strategy. 

Many current approaches for calculating marginal emission rates rely on short-run marginal impacts based on identifying the emission intensity of the marginal generator or using regressions of generation and emissions changes \cite{,daniels2022more,lindberg2022using, sengupta2022current, gagnon2023cambium, spg_mef, gagnon2024consequential}. But are such short-run marginal emissions signals predictive of consequential emissions outcomes when accounting for structural changes in the power system? To answer this question, we estimate charging patterns optimized to minimize short-run marginal emissions and then use our capacity expansion model of the WECC to estimate the consequential emissions impact of this control strategy. To do this, we first estimate the zonal marginal emission rate. Because there is no consensus on how short-run marginal emissions rates from a grid should be computed, we use two approaches. In the first approach (``SRME1''), we assume a uniform 3\% increase in demand for every hour in the target zone and the zonal marginal emission rates are defined as hourly emission increases in WECC normalized by the zonal demand increase. The second approach (``SRME2'') defines marginal emission rates as the total emission increase due to a 1 MW demand increase in the target zone during a specific hour (i.e., demand during other hours remains unchanged). Calculations for each method are detailed in \ref{sec:emisMin}. After quantifying these short-run marginal emission rates, we re-schedule EV charging to minimize emissions using each of these marginal emission rates time series. As marginal emissions cannot be endogenously minimized in a linear optimization model due to nonconvexity\cite{xu2024system}, we iterate the processes described above (quantifying marginal emission rates and minimizing emissions) until consequential emissions from EV adoption almost stay unchanged. All marginal emission rates in the following analysis are from the last iteration. Then we derive new EV dispatch patterns from the last iteration (``Minimize SRME1'' and ``Minimize SRME2'' in Figure \ref{dispatch}) and fix these EV charging demand patterns as inputs to the capacity expansion modeling to capture both structural and operational effects. Finally, we compare the results with the long-run cost optimal case (``Minimize LR-costs''), where EV charge is determined to minimize system costs after considering long-term structural system impacts from increased EV demand. 

\begin{figure}[h]
\centering
\includegraphics[width=1.0\textwidth]{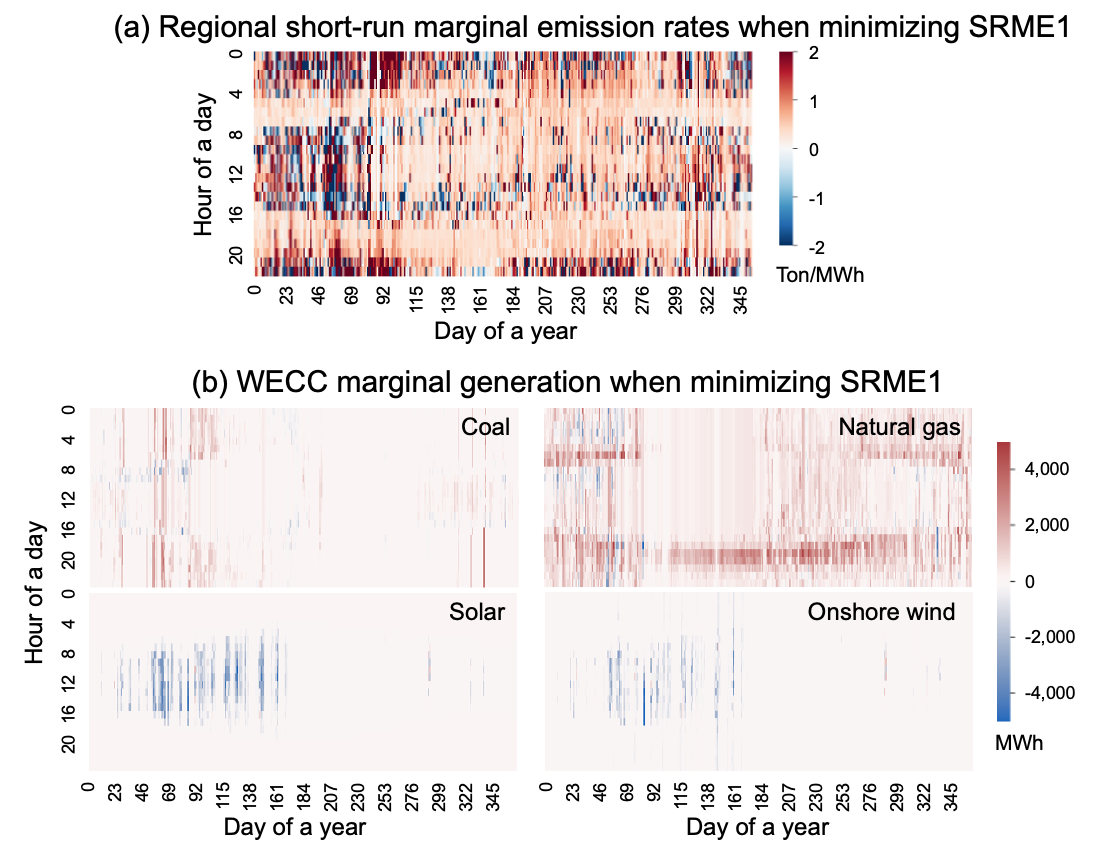}
\caption{Marginal emission rates and generation at the WECC level due to 3\% demand increase in South California (S.CA) under the 24-hour flexibility scenario. The marginal emission rates and generation are after minimizing SRME1. (a) shows the distributions of short-run marginal emission rates and (b) shows marginal generation by technology.}
\label{CA_S}
\end{figure}
\begin{figure}[h]
\centering
\includegraphics[width=1.0\textwidth]{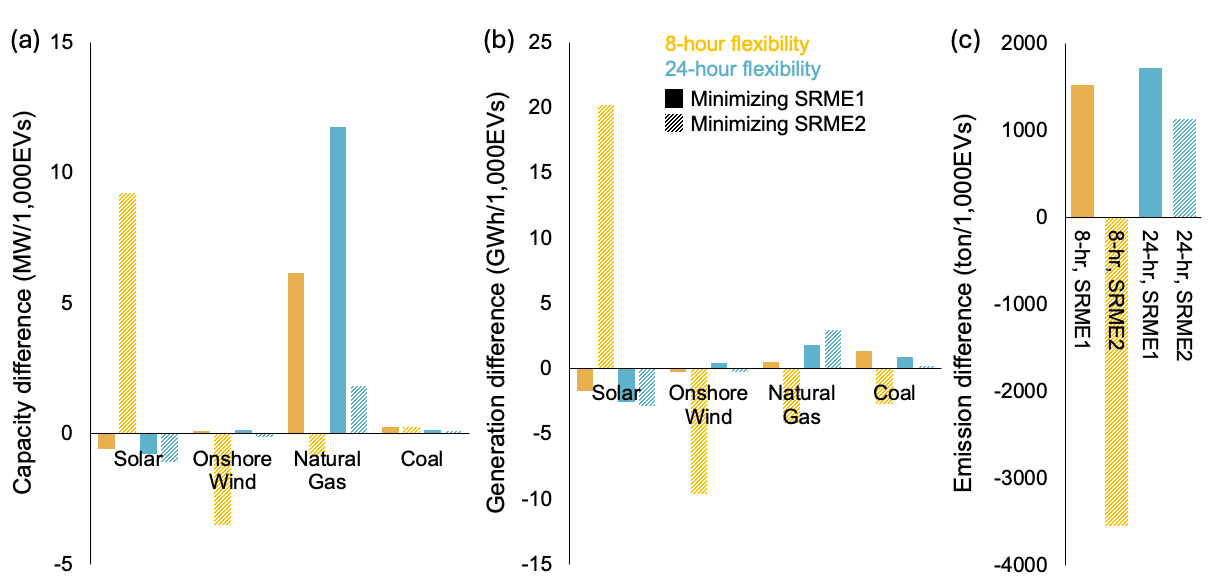}
\caption{Differences between cases minimizing long-run system costs (``LR-costs'') and minimize short-run marginal emissions (``SRME1\&2'') in marginal capacity (a), generation (b), and CO\textsubscript{2} emissions (c) after considering EV charging demand in power system capacity expansion in WECC in 2030 under 8- and 24-hour charging flexibility scenarios. Results are normalized by the number of EVs.}
\label{capacity_flex}
\end{figure}

When calculating marginal emissions based on a uniform increase in demand (the ``SRME1'' case), we observe fossil fuel power plants tend to stay online for more hours, including increased generation during midday. More frequent midday fossil generation occurs because the demand increase in the late afternoon tends to require commitment of more gas- or coal-fired units during midday periods, to ensure that sufficient capacity is available for the evening ramp in demand. This is a result of start-up and shut-down (unit commitment) constraints, which prevent plants from quickly turning on and off during the middle of the day, despite abundant solar supply in these hours. Instead, thermal generators maintain a minimum operational level, driving increased renewable curtailment; thus, higher marginal emission rates may occur during daytime hours, especially during spring and winter (See Figure \ref{CA_S} for a 24-hour flexibility example in South California). Therefore, when SRME1 is used as the demand response signal under both 8- and 24-hour flexibility levels, part of EV charging demand is shifting away from the daytime (when marginal emission rates are high) to late afternoon compared with the ``LR-Costs'' case (Figures S5 and S6). This shift in EV charging patterns impacts the long-term capacity expansion in WECC. Compared with the ``LR-Cost case'', less new solar capacity (-0.6 and -0.8 MW per 1,000 EVs adopted) and more natural gas capacity (6.2 and 11.8 MW per 1,000 EVs adopted) would be added to the power system under the 8- and 24-hour flexibility levels, respectively. Consequently, fossil fuel generation is actually greater when dispatching EVs using the SRME1 signal and total emissions increase by between 1,500 and 1,700 tons per 1,000 EVs added to WECC, compared with the ``LR-Cost case'' (Figure \ref{capacity_flex}). 

\begin{figure}[h]
\centering
\includegraphics[width=1.0\textwidth]{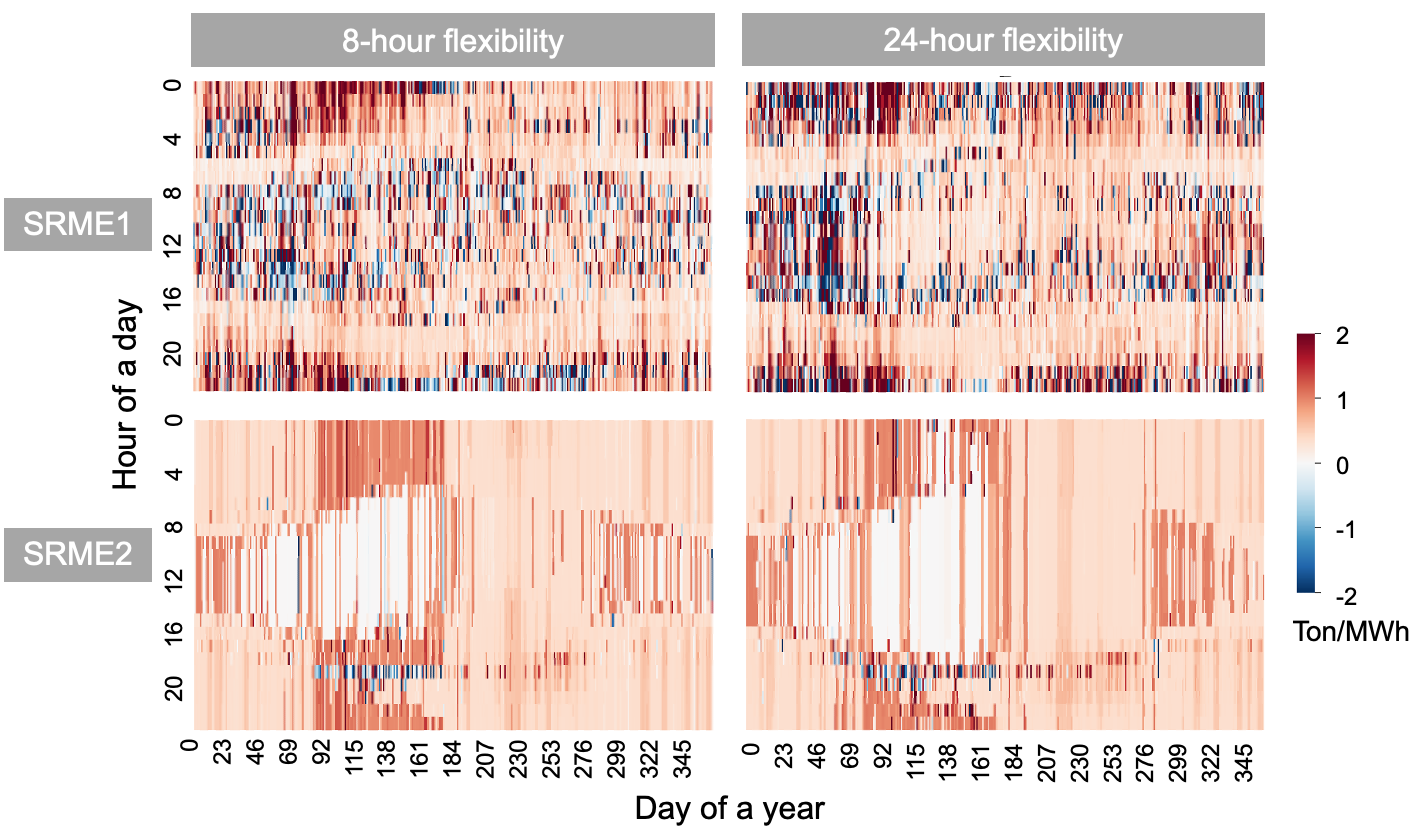}
\caption{Distribution of the short-run marginal emission rates at the WECC level in South California (S.CA). The marginal emission rates are after minimizing short-run marginal emissions. Two different approaches to calculate marginal emission rates are used under ``SRME1'' and ``SRME2''.}
\label{two_srmes}
\end{figure}

When using the second approach, ``SRME2'', which independently increases demand by 1 MW in each zone and in each hour to determine marginal emissions, we observe very different short-run marginal emissions rates from those with SRME1 (see Figure \ref{two_srmes} for an example in South California). SRME2 shows positive emission rates in most periods and negative rates mainly concentrated in the spring and early-summer evenings (this can be explained by additional demand causing gas plants to be committed and stay online due to unit commitment constraints, displacing coal generation over several hours), in both the 8- and 24-hour flexibility cases. This results in a pronounced increase in evening EV demand when EVs are scheduled to minimize SRME2 (Figures \ref{dispatch}, S7, and S8). Scheduling EV charging to minimize emissions using this metric results in different charging patterns between the 8- and 24-hour cases. In the 24-hour flexibility case, mid-day EV charging demand decreases when dispatched to minimize emissions relative to cost-minimizing demand, as illustrated in Figures \ref{dispatch} and S7. This leads to less solar capacity (-1 GW per 1,000 EVs adopted) and more emissions (1,100 ton of CO\textsubscript{2} per 1,000 EVs adopted) in long-run WECC capacity expansion (Figure \ref{capacity_flex}). On the other hand, in the 8-hour case, EV charging that minimizes SRME2 does not show a significant demand decrease in the middle of the day relative to cost-minimizing EV demand (Figures \ref{dispatch} and S8); instead, we observe somewhat increased demand in the early afternoon. Short-run marginal emissions under the 8-hour and 24-hour flexibility cases are different because they are marginal emission rates after minimizing SRME2. The differences in mid-day demand between the 8-hour and 24-hour flexibility cases can be partly explained by the fact that the 8-hour flexibility case exhibits fewer high-emission hours during the winter daytime and more high-emission hours during evening peak-load hours between March and June. The ultimate long-term impact of emissions-based EV scheduling (``Minimizing SRME2'') with 8-hour flexibility is an increase in solar capacity (9 GW per 1,000 EVs adopted) and lower emissions (-3,500 ton of CO\textsubscript{2} per 1,000 EVs adopted). This is completely the opposite effect relative to the 24-hour case or to the case where SRME1 is minimized, illustrating how alternative methods for calculating marginal emissions can produce substantially different operational signals and consequential emissions impacts.

Out of four cases tested in this study (two flexibility levels $\times$ two short-run marginal emission rates), three actually lead to an \textit{increase} in consequential emissions when considering structural changes, despite scheduling charging to minimize short-run emissions. This finding reinforces the limitations of short-run marginal emissions rate metrics as accurate measures of consequential emissions outcomes and raises concerns about the use of these metrics as control-signals for EV dispatch or other scheduling of demand or generation.

\section{Discussion}
In this study, we use capacity expansion models to construct a comprehensive picture of the grid emission impact of EVs. This approach captures not only the short-run effect of additional electricity load, but also any induced structural changes in generation or storage capacity. We find that estimated power system emissions induced by EV charging are three times higher when using short-run marginal emission rate metrics than modeled consequential emissions impacts, because short-run marginal emissions metrics fail to account for the impact of new clean capacity built to meet growing demand. As a result, the climate benefits of EVs are strongly underestimated by short-run emissions metrics. This finding is not expected to be universally applicable, as other regions in the world may find that highly-emitting generators (such as coal power plants) are the resources likely to expand capacity due to EV demand growth. However, in all cases, long- and short-run marginal emissions estimates are likely to fundamentally differ.

Regional average emission rates, which are also frequently used in lifecycle analysis and other estimates of indirect emissions from EV charging, can either under- or over-estimate the true emissions impacts of EV charging, depending on the location of EV adoption. Average emissions rates frequently diverge from consequential emissions outcomes because average emissions are a product of the historical fuel mix, while consequential emissions are primarily a result of \textit{new} capacity additions; unless the two coincidentally align, average emissions rates provide an inaccurate metric to determine consequential emissions from EV adoption. 

Therefore, we suggest considering the structural impacts caused by electrification in different sectors and using long-term metrics capturing induced capacity changes to accurately quantify the climate impacts in life-cycle analysis research. Average emission rates, which are widely used in such studies, often fail to measure climate consequences accurately. Capturing structural changes in capacity is particularly critical for regions with fossil fuel heavy electricity grids that are in the midst of a transition to cleaner energy. For example, several prior studies using average emissions rates to estimate life-cycle emissions from EV adoption have concluded that EVs can lead to greater CO\textsubscript{2} emissions than vehicles consuming fossil fuels in coal-dominated regions of India and China \cite{das2022comparative, hsieh2022integrated, peshin2022should, tang2022life}. However, these analyses often assume a fixed electricity grid and overlook the potential for EV adoption to incentivize new clean energy investments. Such oversights might underestimate the climate benefits of EV adoption, and if these analyses are used to formulate policy, may hinder the development of cleaner transportation in these regions. 

By enabling flexible charging, we also find that EV adoption can achieve greater emission reductions, particularly when 24-hour charging flexibility is available. Therefore, this study contributes to a growing number of studies that highlight the significant emission reduction potential of charging during the daytime in grids where solar PV is a low-cost source of new electricity supply \cite{jochem_assessing_2015, powell_charging_2022,lauvergne_integration_2022}. Achieving this outcome would likely necessitate ubiquitous workplace charging. However, the Edison Electric Institute's projections of EV charging infrastructure estimates that only 17\% of charging in 2030 will occur at the workplace \cite{satterfield_electric_2022}. Instead,  overnight charging incentive programs (similar to the ``8-hr flexibility'' scenario modeled herein) are currently being widely adopted by many utilities. While these programs may save consumers cost, they may not maximize emissions reductions as effectively as daytime charging.

While we find that price-based signals are likely to improve emissions outcomes related to EV charging in many contexts, it should be possible to design control signals for flexible charging that are optimized to reduce emissions impacts. Although prior literature and existing commercial products employ short-run marginal emission rates as a control signal for EV charging\cite{watttime, wang2022feasibility}, we find that scheduling charging to minimize short-run emissions is unlikely to accurately capture the true emissions impacts from EVs and reduce long-term emissions for two reasons. First, while it may appear that short-run marginal emissions metrics reflect empirically observable changes in power system operations, there are actually multiple established methods to estimate this metric, and each can produce substantially different emissions rates. Power system operations are coupled through time (e.g., by unit commitment and storage dispatch decisions), meaning that a change in demand (or supply) in one hour impacts system operations, investment incentives, and emissions outcomes in many other hours. Therefore, short-run marginal emission rate estimates can be highly sensitive to the method used to perturb demand, as well as the assumptions and the temporal and spatial resolution used in the modeling tools. Second, even if we can accurately obtain short-run marginal emissions, they cannot reflect the structural changes that may be induced by anticipated changes in demand patterns and therefore cannot capture the true consequential impacts of EV charging (or other changes in electricity supply and demand). Unfortunately, true consequential emissions outcomes of a given action are fundamentally unobservable, as they depend on changes relative to a counter-factual outcome where the action in question does not occur, and these outcomes are shaped to a large extent by future structural changes in installed capacity. This same challenge has long bedeviled emissions offset markets as well, as true changes relative to a counter-factual are not empirically observable and independently verifiable.\cite{richards2012evaluating, haya2020managing, probst2023systematic} However, it remains theoretically possible to design charging control signals optimized for consequential emissions impacts, if such signals can capture likely structural effects with reasonable accuracy. This remains an important open challenge for future work.  

\section{Acknowledgment}
We acknowledge useful suggestions from Michael Lau in the ZERO Lab at Princeton University when developing the dual-based approach to estimate marginal emission rates; we also acknowledge advising in the initial stages of work from Dr. Qingyu Xu and Professor Denise L. Mauzerall at Princeton University. Funding for this work was provided by the Princeton Zero-carbon Technology Consortium, supported by unrestricted gifts from Google and Breakthrough Energy.

\section{Methods}
\subsection{Model configuration}
To model the power grid in 2030, we use an open-source capacity expansion model, GenX \cite{genx, jenkins2017enhanced}, which is a linear optimization model that determines both electric grid infrastructure investments and operational decisions subject to power system constraints and various cost and policy limitations. The Western Electricity Coordinating Council (WECC) power system is modeled at hourly resolution for a full year (8760 hours in 2030) using an 11-zone system topology based on the EPA's Integrated Planning Model regions \cite{us_epa_documentation_2015} (see Figure S1). These zones roughly correspond to regional transmission constraints, and represent the spatial areas over which demand and generators are aggregated in the model. Existing related policies including state-level Renewable Portfolio Standard (RPS), Clean Energy Standard (CES), and federal-level Production/Investment Tax Credits (PTC/ITC) under the Inflation Reduction Act, are included in the model. Electricity demand profiles, generator parameters, cost scenarios, and other resource inputs to GenX are compiled by the open-source software PowerGenome \cite{powergenome}. More details about input data and GenX setup are included in the supplementary information.

\subsection{Emission impacts estimation}

\begin{figure}[h]
\centering
\includegraphics[width=1.0\textwidth]{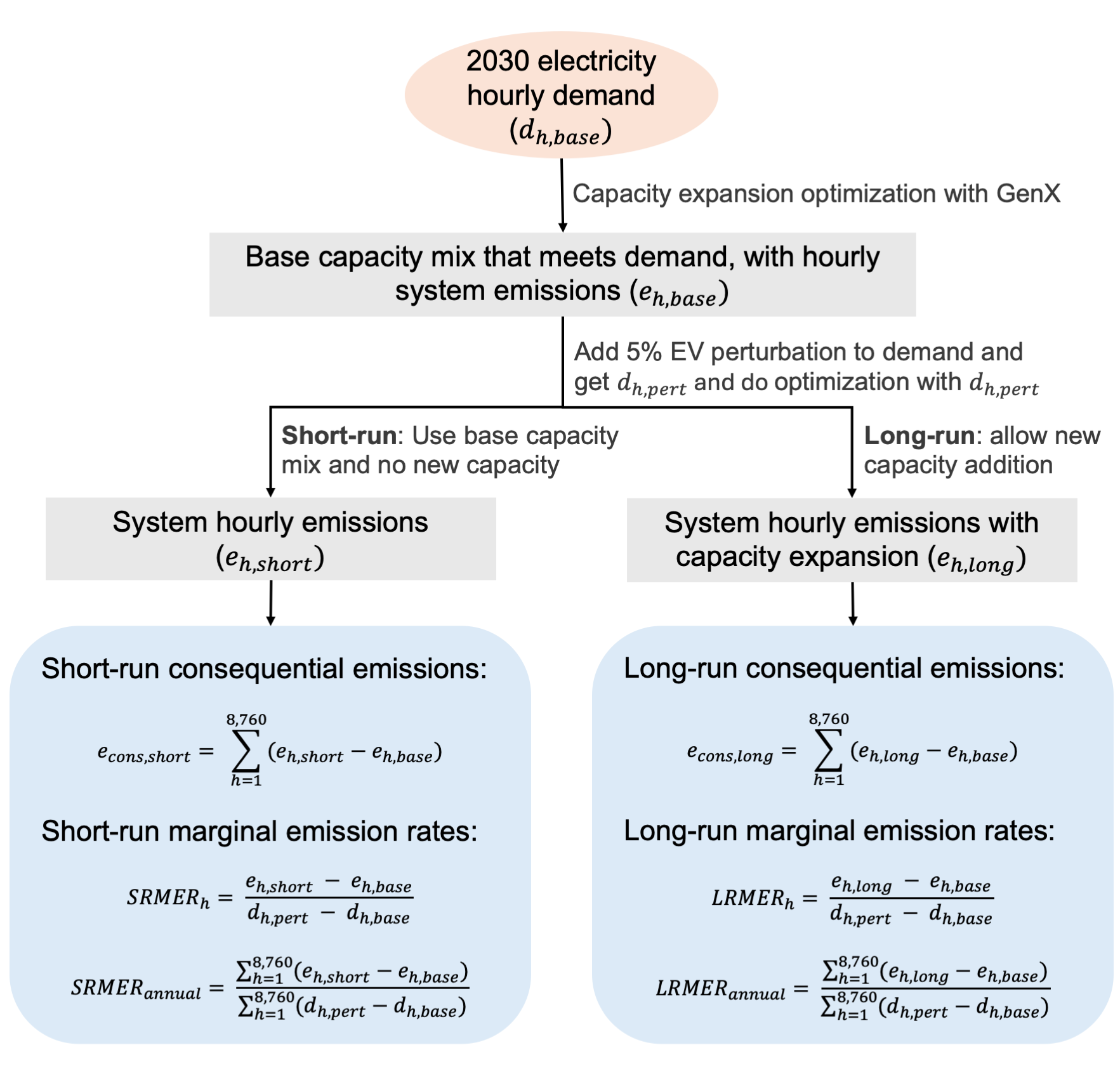}
\caption{Steps to estimate consequential emission impacts from EV adoption. $d_{h, \textrm{base}}$ and $d_{h, \textrm{pert}}$ are the hourly demand at hour $h$ under the base and 5\% EV load perturbed scenarios, respectively. $e_{h, \textrm{base}}$ and $e_{h, \textrm{pert}}$ are the corresponding hourly system emissions.}
\label{fig: emisCal}
\end{figure}

Consequential emission impacts caused by light-duty EV adoption are estimated using short-run and long-run methodologies, respectively (Figure \ref{fig: emisCal}). Marginal emissions are estimate from 5\% perturbation of EV penetration level of the EV stock value under the REPEAT moderate electrification scenario (780,000 EVs)\cite{repeat}. A +5\% perturbation adds about 10 MW of demand during the lowest hour in the EV load profile, and 900 MW during peak demand. For internal combustion engine vehicles, emissions factors were taken from the Argonne National Laboratory's GREET model \cite{greet}, using fuel efficiency projections for 2030 for spark ignition gasoline vehicles.

\subsection{Scenarios design}

To understand how power system marginal emissions vary by EV adoption levels, we include a wide range of EV penetration levels. The base scenario (moderate electrification) for 2030 is projected by the REPEAT project, which has 15.6 million EVs for entire WECC (corresponding to a 20 \% penetration rate). We vary EV penetration levels by $\pm 15\%$ (from a range of 85\% -115\% of the base scenario (or 13.3 million to 17.9 million EVs)). 

We also include two demand flexibility scenarios to reflect EVs' capability of demand response: 8-hour delayed charging (``8-hr flexibility''), and 12-hour advanced and delayed charging (``24-hr flexibility''). The base EV penetration level is used in these demand flexibility scenarios. Since most of the EV charging load is in the evening, the 8-hour flexibility scenario only allows delaying charging to simulate the smart overnight charging programs currently employed by utilities. The 24-hour scenario is intended to demonstrate the potential of significant build-out in workplace charging infrastructure. 

For both base and demand flexibility scenarios, we conduct sensitivity analyses for renewable investment costs and natural gas prices. Different renewable investment costs are obtained from NREL's Annual Technology Baseline 2022 (specifically, NREL ATB's ``conservative'', ``moderate'', and ``advanced'' renewable cost scenarios)\cite{nrel_atb2022}. For natural gas prices, we use EIA's Annual Energy Outlook 2021 Fuel scenarios (specifically the ``reference", ``high resource", and ``low resource" scenarios)\cite{aeo2021}. Moderate renewable costs and reference natural gas prices are used in the main analyses.

We also investigate the impact of EV adoption in each zone of WECC, to understand how regional variation in infrastructure could affect EV-related emissions. We increase the EV penetration level by 5\% in each zone compared to the base case, keep the EV demand unchanged in other zones, repeat the steps listed in Figure \ref{fig: emisCal} for each zone, and calculate system emission impacts associated with zonal EV adoption.

\subsection{Minimize short-run marginal emissions}\label{sec:emisMin}
To understand emission impacts from using marginal emission rates as a signal to guide EV charging, we estimate the hourly ($h$) marginal emission rates at the zonal level ($z$) by two approaches. For both approaches, we run a base operational case in GenX with fixed power plants and transmission lines capacity, where we remove EV charging demand. Then, in the first approach (``SRME1''), we increase electricity load by 3\% in the target zone for every hour in 2030, determine power system operations with GenX, quantify the hourly emission increases in WECC compared with those from the base case ($\Delta Emissions\textsubscript{h,z}$), and calculate the marginal emission rates at hour $h$ in zone $z$ as: $\frac{\Delta Emissions\textsubscript{h,z}}{3\% \times Annual Zonal Load\textsubscript{z}}$. We repeat the calculation for each zone in WECC and get the marginal emission rates that vary by zone. In the second approach (``SRME2''), we use duality theory to estimate the emissions change from a 1-MW change in EV demand in a given hour and zone. This method involves the following steps. First, we obtain the objective value from the base case (system costs, $obj\textsubscript{0}$) and the dual value of the demand constraint in a given hour and zone ($\mu^{base}_{h,z}$). Then, in a second run, we add a new constraint to GenX where we limit the system costs to be less than or equal to $obj\textsubscript{0}$ (let the dual of this constraint be $\lambda^{second}$) and minimize system emissions, instead of minimizing costs. The dual values of the demand constraint in the second run ($\mu^{second}_{h,z}$) allow us to estimate locational marginal emission rates using the formula: $\mu^{second}_{h,z} - \lambda^{second} \times \mu^{base}_{h,z}$. The subtraction of the second term (which represents the emissions impact of the newly-introduced constraint enforcing system costs to be at their minimum value $obj\textsubscript{0}$) is necessary to isolate the impact of demand on emissions. A detailed description of this approach is provided in the supplementary information.

After obtaining these emission rates, we reformulate GenX to minimize marginal emissions when charging EVs. However, as discussed in Xu et al. (2024), SRME cannot be endogenously minimized in a linear capacity expansion model due to nonconvexity\cite{xu2024system}. Therefore, we use results from the short-run flexibility scenarios as the starting baseline (B\textsubscript{0})and calculate the first set of hourly zonal SRMERs (R\textsubscript{0}). To minimize total marginal emissions from the system, we include a high emission penalty cost (\$ 1,000 per ton) for marginal emissions in the objective function of GenX and re-do the optimization similar to the base flexible case (using the original demand curve and allowing EV flexible charging) to get a new base case (B\textsubscript{1}), which is used to calculate R\textsubscript{1}. We repeat the process until the emission impacts from EV adoption ``almost" converge: differences in consequential emissions from 5\% EV adoption increase are less than 1\% (B\textsubscript{n}).

To understand the long-term impacts of EV charging pattern designed to minimize short-run electricity prices and marginal emissions, we fix the charging demand profile from B\textsubscript{0} and B\textsubscript{n} and do capacity expansion for WECC in GenX, respectively. Then we compare the impacts of using short-run marginal emission rates as demand response signals with those from long-run flexibility scenarios (LR-costs).

\newpage
\printbibliography 
\end{document}


\title{Are EVs Cleaner Than We Think? Evaluating Consequential Greenhouse Gas Emissions from EV Charging}

\author[1]{\fnm{Riti} \sur{Bhandarkar}}\email{ritib@princeton.edu}
\equalcont{These authors contributed equally to this work.}
\author*[1]{\fnm{Qian} \sur{Luo}}\email{ql7299@princeton.edu}
\equalcont{These authors contributed equally to this work.}
\author[1]{\fnm{Emil} \sur{Dimanchev}}\email{ed0400@princeton.edu}
 
\author[1,2]{\fnm{Jesse} \sur{D. Jenkins}}\email{jdj2@princeton.edu}

\affil[1]{\orgdiv{Andlinger Center for Energy and the Environment}, \orgname{Princeton University}, \orgaddress{\city{Princeton}, \postcode{08540}, \state{NJ}, \country{USA}}}
\affil[2]{\orgdiv{Department of Mechanical and Aerospace Engineering }, \orgname{Princeton University}, \orgaddress{ \city{Princeton}, \postcode{08540}, \state{NJ}, \country{USA}}}

\maketitle

\section{GenX Setup}

\begin{figure}[h]
\includegraphics[scale = 0.4]{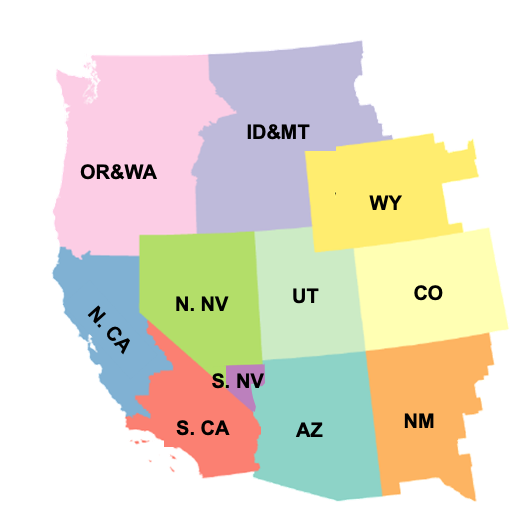}
\centering
\caption{Map of the 11-zone configuration of WECC system.}
\label{wecc-map}
\end{figure}

The following input data parameters are used in GenX:
\begin{itemize}
\item New capacity build was permitted for the following technologies: solar, onshore and offshore wind, natural gas (combined cycle and combustion turbine), battery storage, and geothermal. Technologies not permitted to build new capacity were conventional hydroelectricity, hydro pumped storage, biomass, and coal
\item The 2.6 cents per KWh PTC credit for wind and solar from the Inflation Reduction Act were implemented into the model as a subsidy distributed over the lifetime of the project, with a weighted average cost of capital of 3.8\% for wind and 2.6\% for solar. ITC credits of 30\% were implemented for offshore wind and battery resources.
\item Base case technology costs were obtained from NREL's Annual Technology Baseline 2022 'moderate' scenario \cite{nrel_national_renewable_energy_laboratory_2022_2022}, and generator heat rates were obtained from the EIA's AEO 2019 Cost and Performance Report \cite{us_energy_information_administration_eia_annual_2021}. Fuel costs were from the "reference" scenario. 
\item Electricity demand profiles were based on projections from the NREL Electrification Futures Study study and the Princeton University REPEAT project for flexible demand technology stock numbers\cite{mai_electrification_2018, repeat}. 
\end{itemize}


\begin{table}
 \caption{\label{table_gendata} Cost and capacity data for the technologies modeled in the capacity expansion modeling runs.}
\begin{tabular}{ |p{5cm}|p{2.5cm}|p{2cm}|p{2.5cm}|  }
 \hline
 \multicolumn{4}{|c|}{\textbf{Generator Data Parameters}} \\
 \hline
 \textbf{Technology }&\textbf{Fixed O\&M (\$/MWh-yr)}&\textbf{Variable O\&M (\$/MWh)} &\textbf{Existing Capacity (2020) (MW)}\\
 \hline
 Battery Storage &   0  & 0   &248\\
 Biomass   & 149,833    &0   &655.2\\
 Coal & 60,000-75,000 & 1.88 &12,770.8 \\
 Hydroelectric &47,048 & 0&  49,169.2\\
 Distributed Generation &n/a &n/a &18,896\\
 Geothermal &101451 - 209101 & 0 &1,957.8\\
 Hydroelectric Pumped Storage & 47,048
 & 0 & 4762.2 \\
 Natural Gas Combined Cycle & 11,144 - 16,492 & 2-4.55 &57,378.2\\
 Natural Gas Combustion Turbine &12,374-14,524 & 5-5.23 &26,293.7\\
 Natural Gas Steam Turbine & 36,367-51,313& 1.06 &4,980.3 \\
 Nuclear & 193,275-275,171 & 0 &5409.6 \\
 Offshore Wind & 69,365 &0 &0 \\
 Onshore Wind &40,367-43,000
 & -13.08-0 &31,872.7 \\
 Solar Photovolatic &17,282-22,887 & -12.28-0
&35850.7  \\
 
\hline
\end{tabular}
\end{table}


\section{Dual-based marginal emission rates}

In this section, we describe a dual-based approach for calculating short-run marginal emission rates, which is used in the ``SRME2" case in the main text. This method estimates the change in system-wide annual emissions resulting from a 1 MW increase in demand in a given zone $z$ at a given time $t$. We denote this short-run marginal emissions rate as $\epsilon_{z,t}$. At a high level, the approach entails running a cost-minimizing power system model in a first step, and, in a second step, running a modified power system model that minimizes emissions, which allows the derivation of $\epsilon_{z,t}$ using duality theory\footnote{It is worth noting that a version of this method can be used for the estimation of marginal impacts other than emissions, depending on the objective function used in the modified problem, \eqref{model_co2min}.}. 

First, consider the following abstract representation of a linear power system dispatch model, with a classical power balance constraint: \eqref{eq_powerbal}, and physical constraints on generation, storage, and transmission flows, such as capacity limits: \eqref{eq_constr_misc}.

\begin{subequations} \label{model_costmin}
\begin{alignat}{4}
& \min_{x} \; \;  && C(x) \\
& \text{s.t.} &&  \mathbf{1}x = \overline{D} \quad &(\mu^{base}) \label{eq_powerbal}\\
&&& Ax \leq b \label{eq_constr_misc}
\end{alignat}
\end{subequations}

Where,
\begin{itemize}
    \item $x$: Power supply (generation, storage, transmission flows; storage charging and transmission outflows are represented with negative values).
    \item $C(x)$: Total system cost.
    \item $\overline{D}$: Demand level.
    \item $A$, $b$: Parameters for physical constraints.
\end{itemize}

Note that the dual variable $\mu^{base}$ represents the \textbf{marginal cost of electricity}.

The first step of our method is to solve model \eqref{model_costmin} and obtain the optimal objective function value, denoted $\overline{C}$, and marginal cost of electricity, $\mu^{base}.$

Secondly, consider the modified power system model \eqref{model_co2min} below, which introduces two changes: first, it minimizes emissions, denoted $E(x)$, and second, it contains a new constraint, \eqref{eq_costconstr}, which ensures that total costs do not exceed the the minimum cost level $\overline{C}$ derived from the optimal objective function value of \eqref{model_costmin}. The purpose of changing the objective function in \eqref{model_co2min} is not to reduce emissions beyond what would occur in a least-cost solution. Instead, this is merely meant to allow us to derive information about the short-run marginal emissions of demand by using the dual $\mu^{second}$, as will be shown below.

\begin{subequations} \label{model_co2min} 
\begin{alignat}{4}
& \min_{x} \; \;  && E(x) \\
& \text{s.t.} &&  \mathbf{1}x = \overline{D} \quad & (\mu^{second}) \label{eq_powerbal2} \\
&&& Ax \leq b &(\gamma) \\
&&& \overline{C} \geq C(x) \quad & (\lambda^{second}) \label{eq_costconstr}
\end{alignat}
\end{subequations}

Note that the modified problem \eqref{model_co2min} captures least-cost power system decisions, as formalized in Proposition \eqref{prop_1}. 
\vspace{10px}
\begin{prop} \label{prop_1}
    \eqref{model_co2min} $\implies$ \eqref{model_costmin}; i.e., any solution to \eqref{model_co2min} also solves \eqref{model_costmin}. 
\end{prop}
\vspace{-10px}
\begin{proof}
Any solution to \eqref{model_co2min} is feasible for \eqref{model_costmin}, as \eqref{model_co2min} has the same feasible region as \eqref{model_costmin}. To see this, observe that the new constraint \eqref{eq_costconstr} does not change the feasible region of \eqref{model_costmin}. Moreover, any solution to \eqref{model_co2min} optimizes \eqref{model_costmin} because \eqref{eq_costconstr} constrains the objective of \eqref{model_costmin} to its optimal value. Note that $C(x) = \overline{C}$ at the optimal solution of \eqref{model_co2min}, since $C(x) < \overline{C}$ is infeasible for \eqref{model_costmin}, and therefore infeasible for \eqref{model_co2min}.
\end{proof}

\begin{remark}
    A solution to \eqref{model_costmin} is not necessarily a solution to \eqref{model_co2min}, because of the possibility of multiple least-cost solutions with different emission levels. However, this does not affect the applicability of our method, which merely relies on \eqref{model_co2min} providing a least-cost solution.
\end{remark}

Solving problem \eqref{model_co2min} allows us to estimate $\epsilon_{z,t}$ using the following formula:
\vspace{10px}
\begin{prop}
    $\epsilon_{z,t} = \mu^{second}_{z,t} - \lambda^{second} \mu^{base}_{z,t}$; i.e., the short-run marginal emissions $\epsilon_{z,t}$ are a function of the dual of the power balance constraint \eqref{eq_powerbal2} and the dual of the least-cost constraint \eqref{eq_costconstr}.
\end{prop}
\vspace{-10px}
\begin{proof}
The Lagrangian function for \eqref{model_co2min} is defined as:
\begin{alignat}{3}
    &\mathcal{L}(x, \lambda^{second}, \mu^{second}, \gamma) = && E(x)  - \mu^{second\;T}(x - \overline{D}) \notag \\
    &&&+ \gamma^T(Ax-b) - \lambda^{second\;T} (\overline{C} - C(x))  \label{eq_lagrang}
\end{alignat}

The following stationarity condition holds at the optimal solution of \eqref{model_co2min} with respect to supply from the marginal technology in zone $z$, at time $t$, which we denote $x_{m,z,t}$:
\begin{equation}
\frac{\partial \mathcal{L}}{\partial x_{m,z,t}} = \frac{\partial E}{\partial x_{m,z,t}} - \mu^{second}_{z,t} + A\gamma_{m,z,t} + \lambda^{second} \frac{\partial C}{\partial x_{m,z,t}}  = 0 \label{eq_stationarity}
\end{equation}

Where $\frac{\partial E}{\partial x_{m,z,t}} $ represents the emissions intensity of the marginal technology $m$, and $A\gamma_{m,z,t}$ represents the impact of each of the physical supply constraints on emissions. 

Rearranging:

\begin{equation}
\frac{\partial E}{\partial x_{m,z,t}} + A\gamma_{m,z,t} = \mu^{second}_{z,t} - \lambda^{second} \frac{\partial C}{\partial x_{m,z,t}}  \label{eq_stationarity_rearranged}
\end{equation}

Next, recall that $\frac{\partial C}{\partial x_{m,z,t}}$ represents the marginal cost of electricity, which we already obtained from the dual solution of model \eqref{model_costmin}:
\begin{equation}
\frac{\partial C}{\partial x_{m,z,t}} = \mu^{base}_{z,t} \label{eq_mu_base}
\end{equation}

Finally, note that $\frac{\partial E}{\partial x_{m,z,t}} + A\gamma_{m,z,t}$ represents the total impact of marginal supply on emissions, capturing both the emissions intensity of generation (the first term) and the influence of all physical constraints (the second term); thus: 

\begin{equation}
\frac{\partial E}{\partial x_{m,z,t}} + A\gamma_{m,z,t} = \epsilon_{z,t} \label{eq_epsilon}
\end{equation}

Substituting \eqref{eq_mu_base} and \eqref{eq_epsilon} into \eqref{eq_stationarity_rearranged}:

\begin{equation}
    \epsilon_{z,t} = \mu^{second}_{z,t} - \lambda^{second} \mu^{base}_{z,t} 
\end{equation}
\end{proof}

To build intuition about Proposition 2, consider first a simple case with only one zone and no storage, where the only constraints in $Ax \leq b$ are maximum capacity constraints on generation. In this case, it would be expected that the marginal emissions would equal the emissions intensity of the marginal generator. Indeed, the term $A\gamma$ vanishes in this case because $\gamma=0$. This is because, by construction, marginal demand has to be met by a resource that is not capacity constrained. Specifically, note that, $x_{m,t}$ would represent either generation or load shedding. In the former case capacity limits do not bind, i.e.: $Ax < b$, implying $\gamma_{m,t} = 0$ by complementary slackness. In all other periods, $x_m$ represents load shedding, which we assume to be unconstrained, leading to $\gamma_{m,t} = 0$. Thus Proposition 2 means that $\mu^{second}_{t} - \lambda^{second} \mu^{base}_{t}$ would equal the emissions intensity of the marginal generator, $\frac{\partial E}{\partial x_{m,t}}$. 

In more realistic cases, $A\gamma_{m,z,t}$  captures the emissions impacts of physical constraints, such as the impact of storage constraints on emissions. For example, when marginal demand would cause a reduction in storage charging, $\gamma$ captures the emissions impact (i.e., the generation emissions that would be caused at a later period by the absence of the stored energy), and $A$ captures the round-trip efficiency of the storage technology. 

\section{Additional Results}

\begin{figure}[h]
\centering
\includegraphics[width=1.0\textwidth]{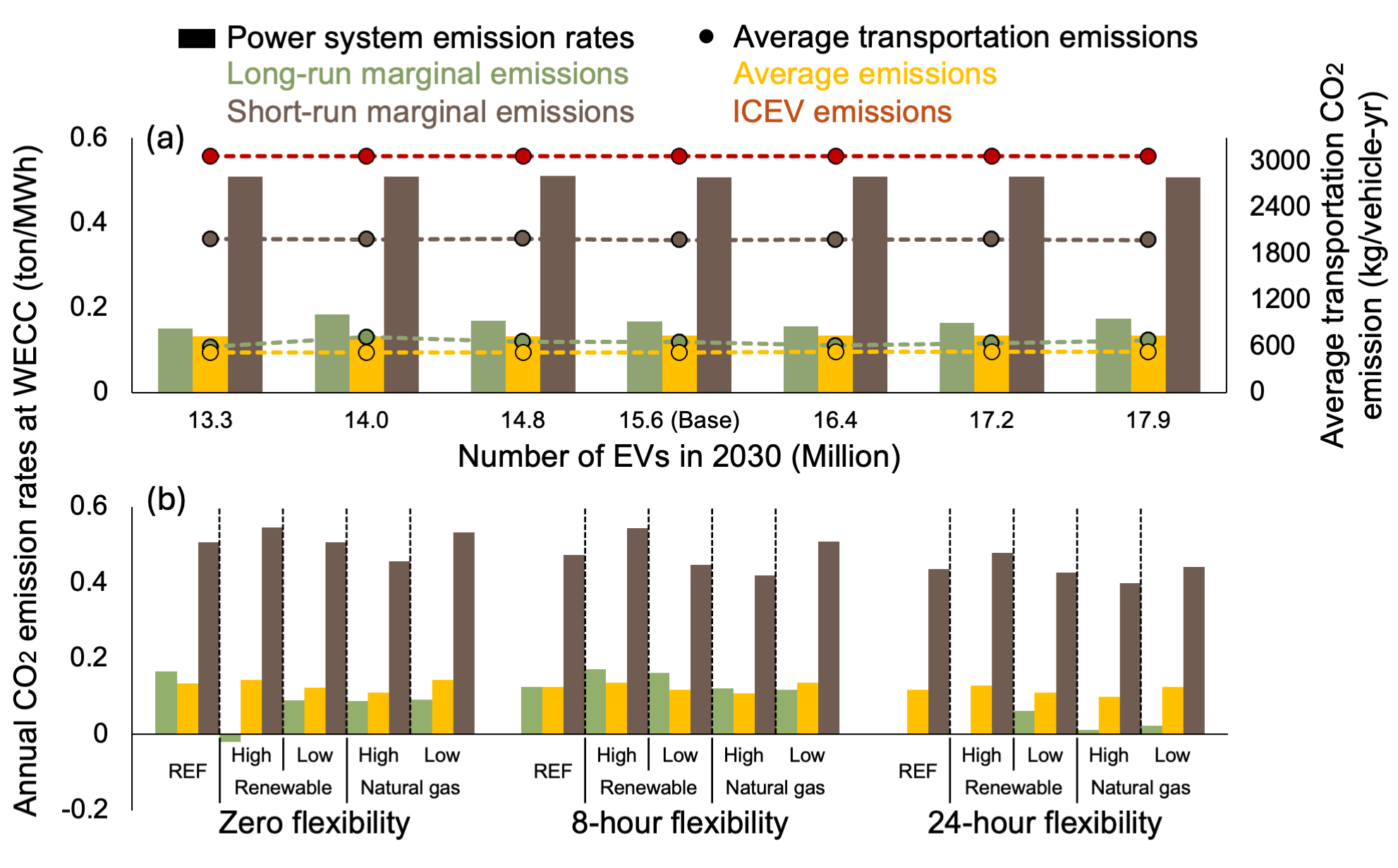}
\caption{Annual long-run, short-run marginal, and average WECC power system emissions varying by EV adoption levels (a), by EV locations (b), and by renewable investment costs and natural gas prices (c). Results in (b) and (c) are based on the ``Base" EV stock numbers (15.6 million EVs).}
\label{LRSRAve_sensitivity}
\end{figure}

\begin{figure}[h]
\centering
\includegraphics[width=1.0\textwidth]{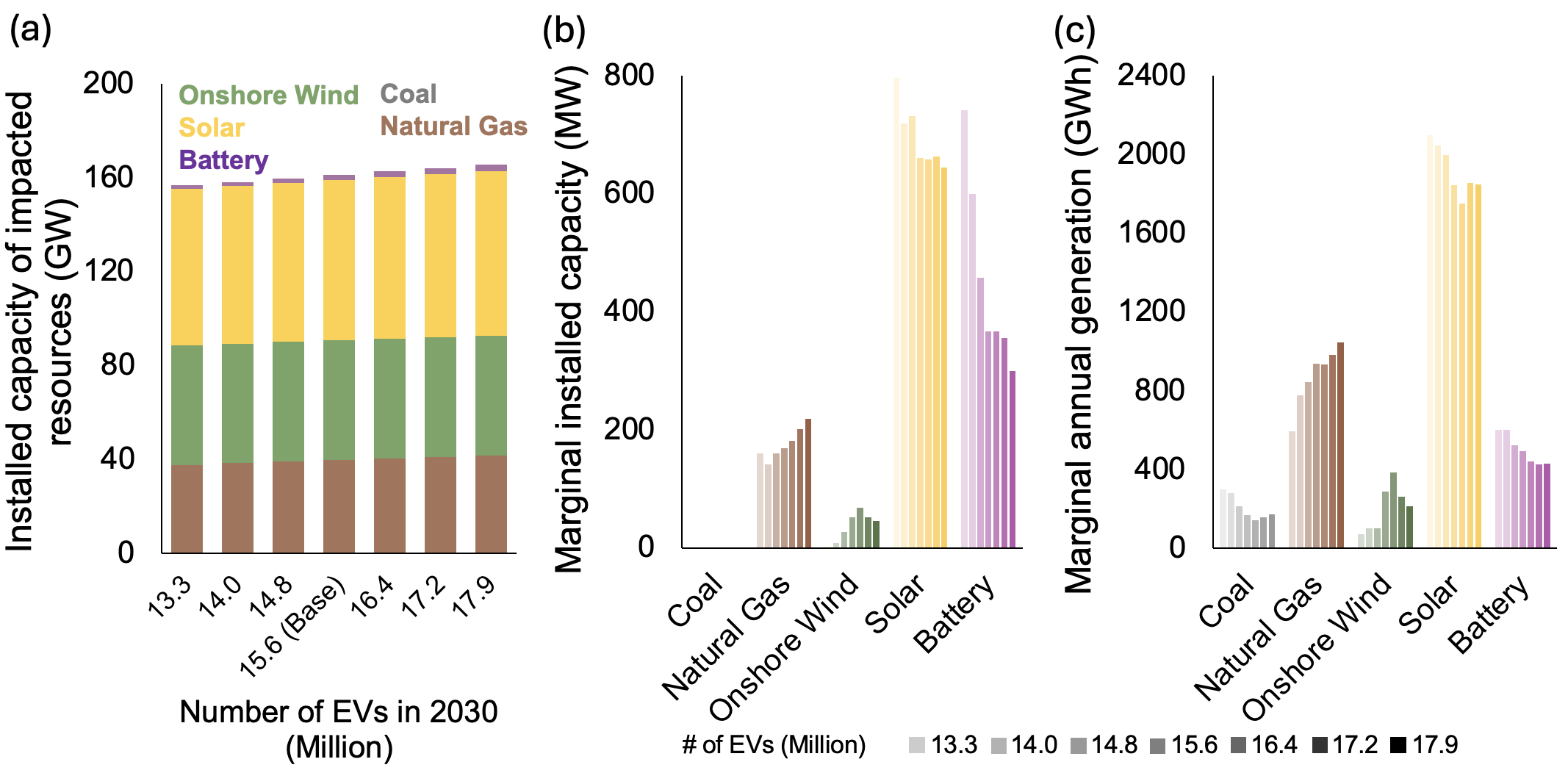}
\caption{Total installed capacity (a), marginal installed capacity (b), and marginal annual generation (c) of energy resources that are impacted by 5\% EV demand increase (0.78 million EVs) under various EV adoption level. 15.6 million EV is the base scenario, assuming that the EV penetration level is 20\% in 2030. }
\label{capacity_0flex}
\end{figure}

\begin{figure}[h]
\centering
\includegraphics[width=1.0\textwidth]{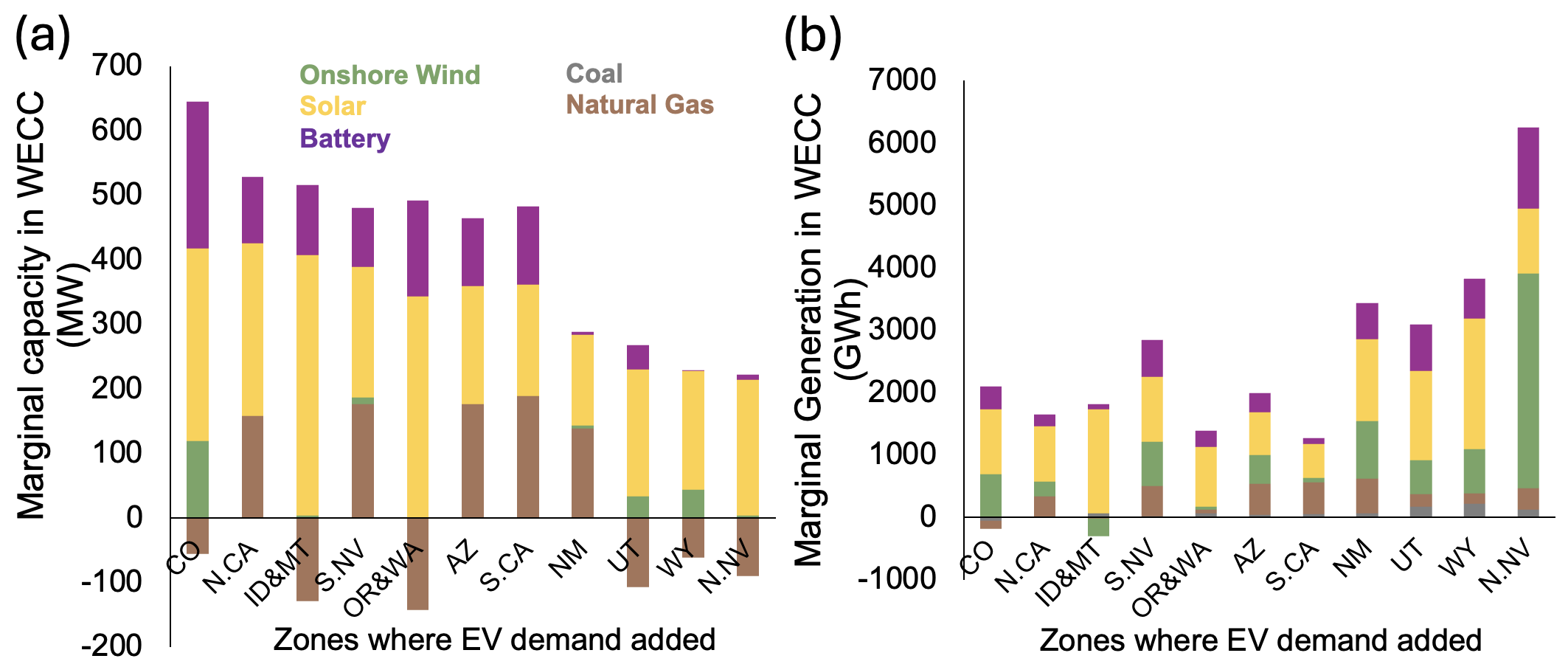}
\caption{WECC system-level marginal installed capacity (b), and marginal annual generation (b) of energy resources that are impacted by 5\% EV demand increase in each zone. Marginal capacity and generation are normalized by added EV demand (1TWh of demand increase).}
\label{capacity_0flex_zonal}
\end{figure}

\begin{figure}[h]
\centering
\includegraphics[width=1.0\textwidth]{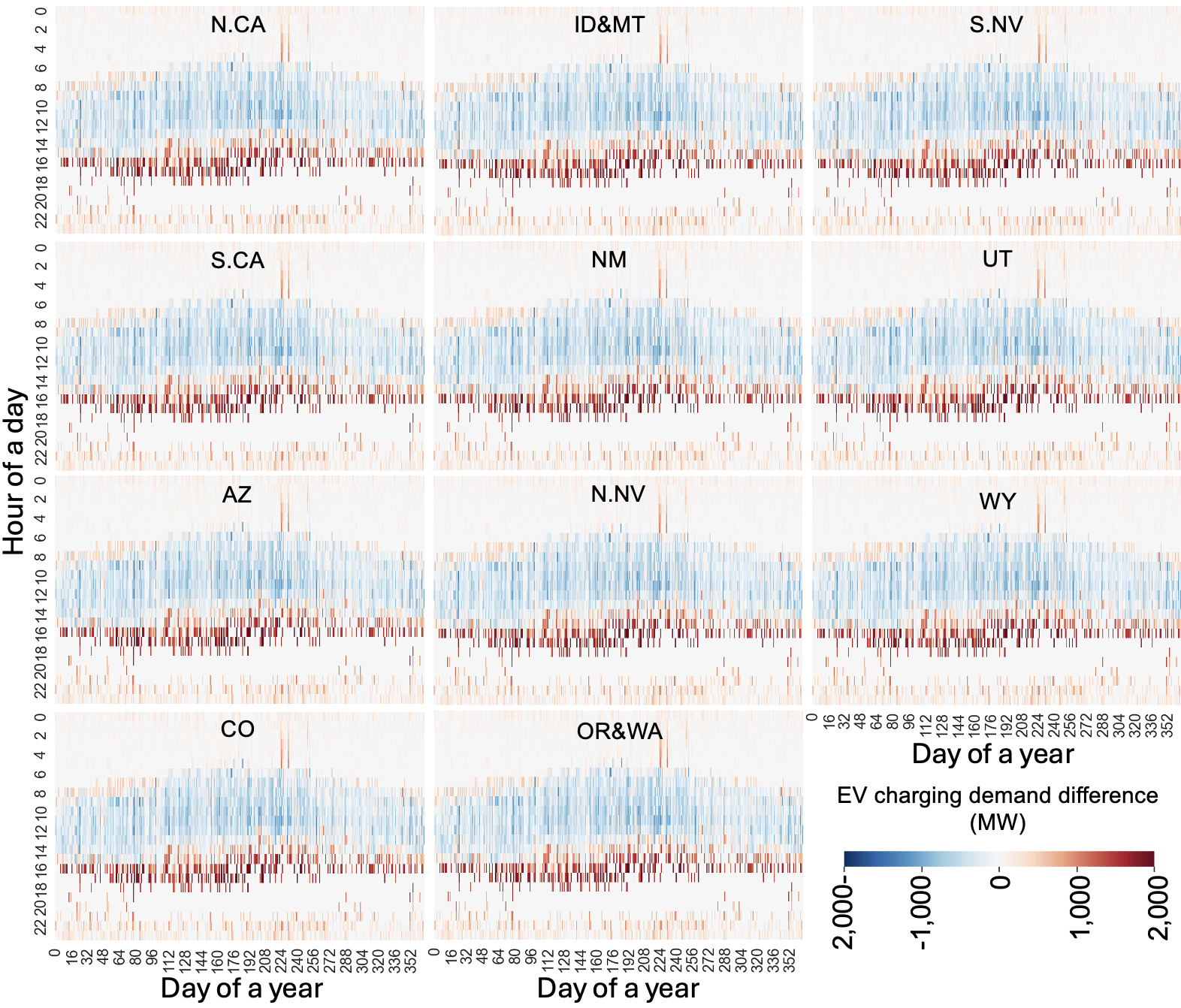}
\caption{Hourly distribution of EV charging demand differences between the short-run marginal emission minimization (``Minimize SRME1'') and long-run cost minimization cases (``Minimize LR-Costs'') in each zone of WECC with 24-hour flexibility levels.}
\label{demandByZone_24hr_srme1}
\end{figure}

\begin{figure}[h]
\centering
\includegraphics[width=1.0\textwidth]{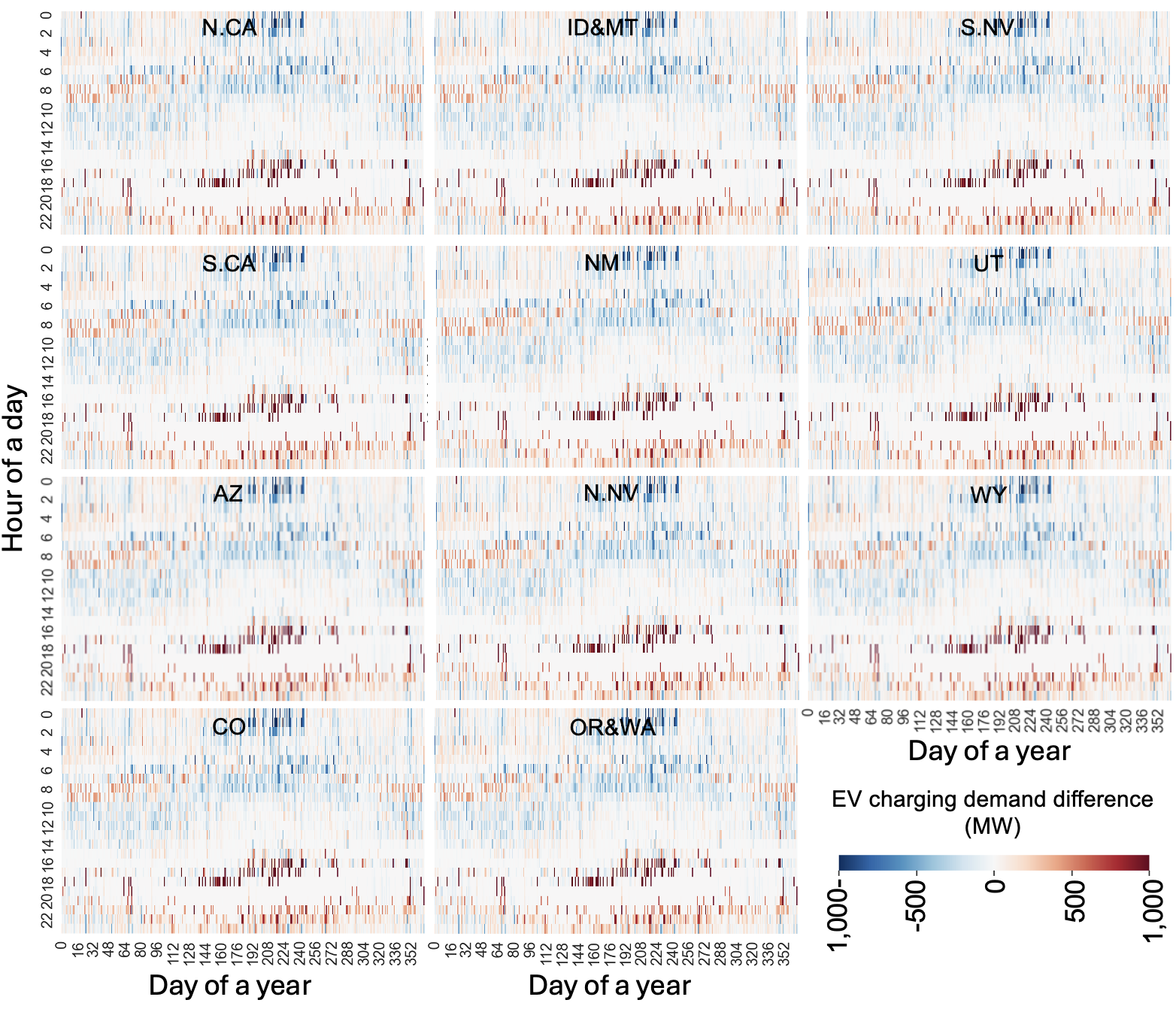}
\caption{Hourly distribution of EV charging demand differences between the short-run marginal emission minimization (``Minimize SRME1'') and long-run cost minimization cases (``Minimize LR-Costs'') in each zone of WECC with 8-hour flexibility levels.}
\label{demandByZone_8hr_srme1}
\end{figure}

\begin{figure}[h]
\centering
\includegraphics[width=1.0\textwidth]{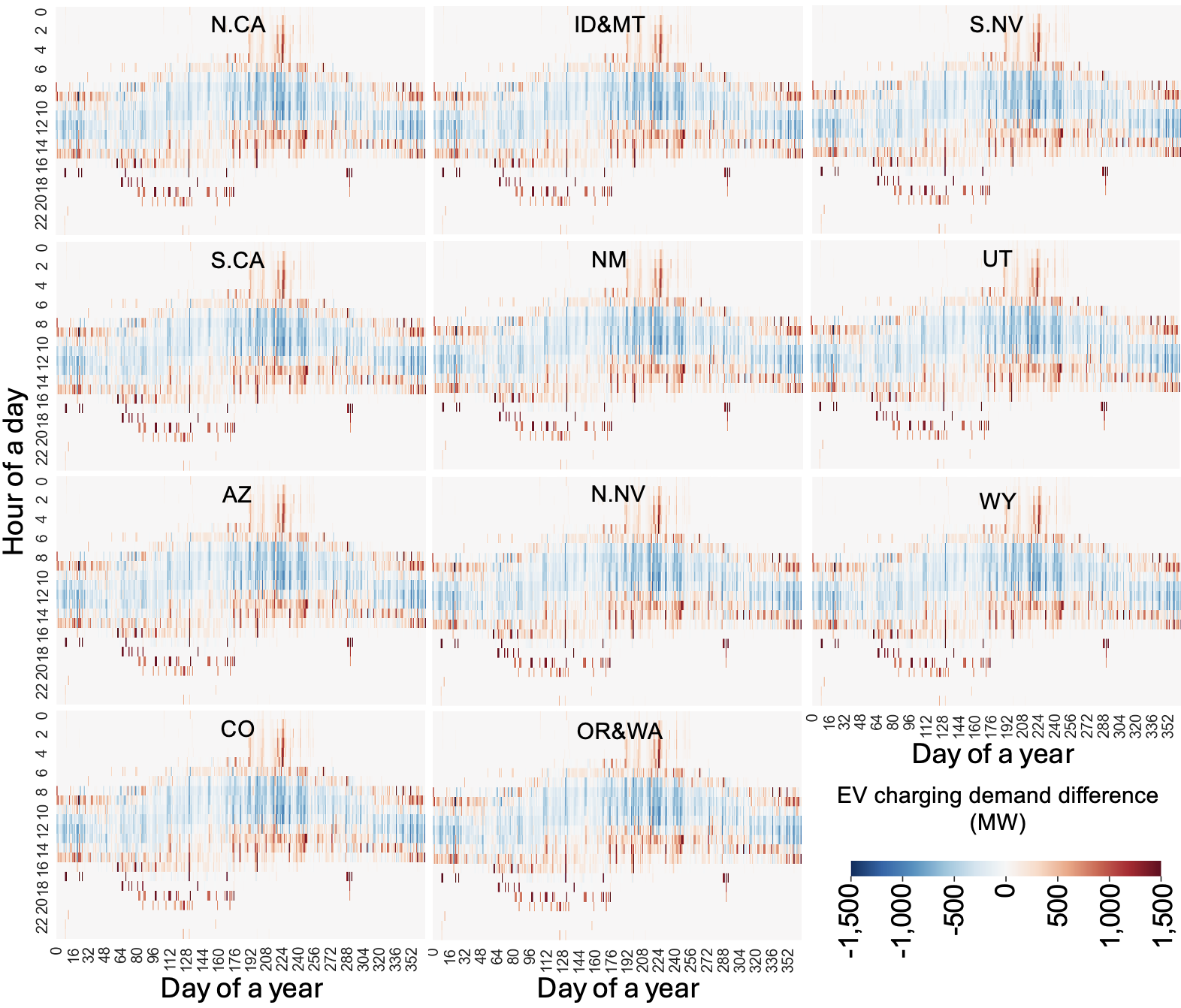}
\caption{Hourly distribution of EV charging demand differences between the short-run marginal emission minimization (``Minimize SRME2'') and long-run cost minimization cases (``Minimize LR-Costs'') in each zone of WECC with 24-hour flexibility levels.}
\label{demandByZone_24hr_srme2}
\end{figure}

\begin{figure}[h]
\centering
\includegraphics[width=1.0\textwidth]{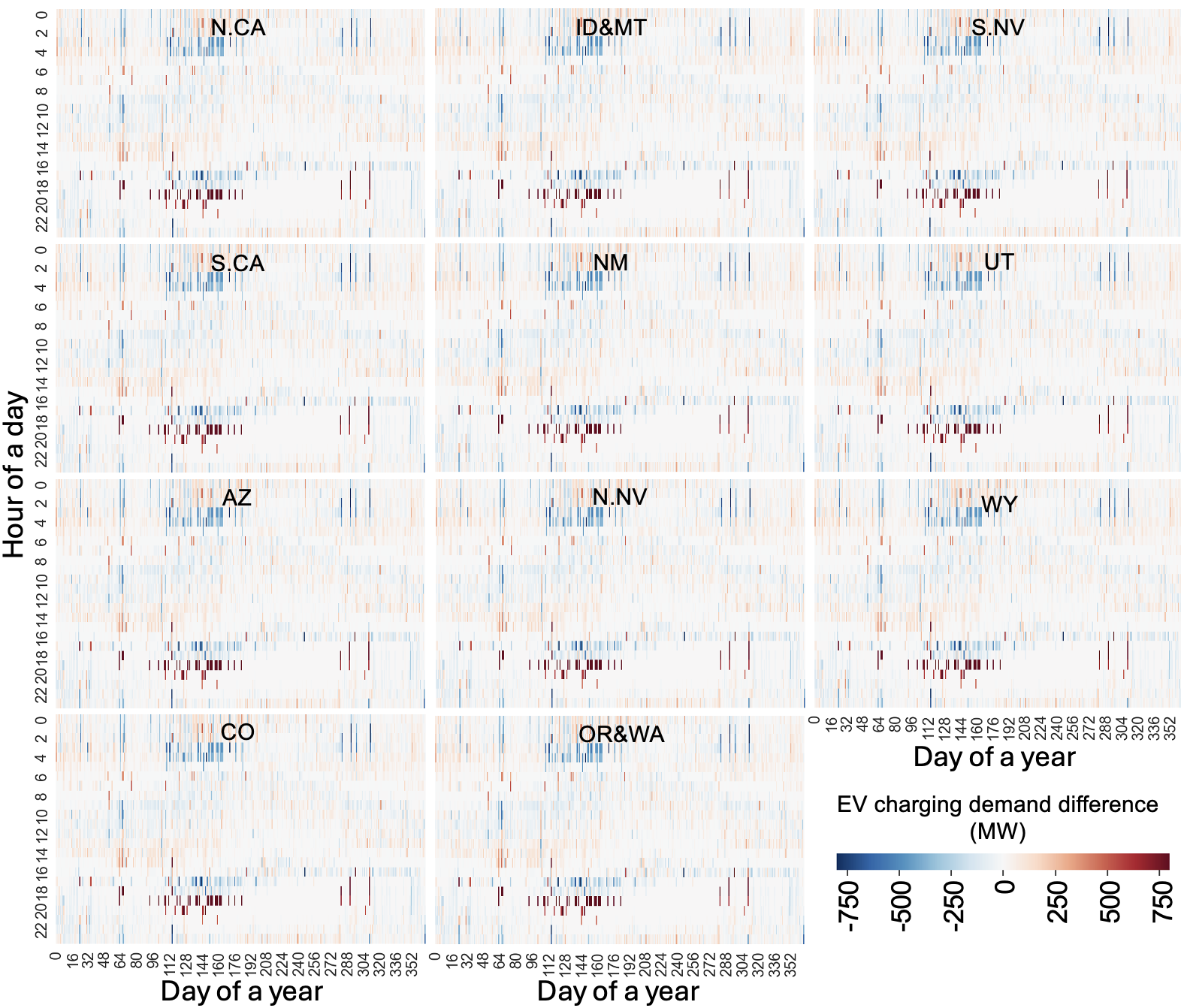}
\caption{Hourly distribution of EV charging demand differences between the short-run marginal emission minimization (``Minimize SRME2'') and long-run cost minimization cases (``Minimize LR-Costs'') in each zone of WECC with 8-hour flexibility levels.}
\label{demandByZone_8hr_srme2}
\end{figure}

\clearpage
\printbibliography 